\documentclass[12pt]{article}
\usepackage{a4wide}
\usepackage{latexsym}
\usepackage{amsfonts}
\usepackage{amssymb}
\usepackage{axodraw}

\def\l{\langle}
\def\r{\rangle} 
\def\bq{\begin{eqnarray}}
\def\eq{\end{eqnarray}}

\def\bs{\begin{small}}
\def\es{\end{small}}
\def\eps{\varepsilon}

\def\topoA{A}

\def\topoC{C}
\def\topoD{D}
\def\topoE{E}

\def\topoa{a}
\def\topob{b}
\def\topoc{c}
\def\topod{d}
\def\topoe{e}
\def\topof{f}
\def\topog{g}
\def\topoh{h}

\def\onedot(#1){
 \begin{picture}(35,10)(0,5)
  \Vertex(5,5){2}
  \Text(9,5)[l]{$#1$}
 \end{picture} 
}
\def\twodot(#1,#2){
 \begin{picture}(35,30)(-5,10)
  \Vertex(5,5){2}
  \Vertex(5,25){2}
  \Line(5,5)(5,25)
  \Text(9,25)[l]{$#1$}
  \Text(9,5)[l]{$#2$}
 \end{picture} 
}

\def\spa#1.#2{\left\langle#1\,#2\right\rangle}
\def\spb#1.#2{\left[#1\,#2\right]}
\def\spab#1.#2.#3{\langle\mskip-1mu{#1} 
                  | #2 | {#3}\mskip-1mu\rangle}

\begin{document}

\thispagestyle{empty}

\begin{flushright}
UPRF-03-002
\\
\end{flushright}

\vspace{1.5cm}

\begin{center}
  {\Large \bf Subtraction terms at NNLO}\\[.3cm]
  \vspace{1.7cm}
  {\sc Stefan Weinzierl$^{1}$}\\
  \vspace{1cm}
  {\it Dipartimento di Fisica, Universit\`a di Parma,\\
       INFN Gruppo Collegato di Parma, 43100 Parma, Italy} \\
\end{center}

\vspace{2cm}

% abstract ---------------------------------------
\begin{abstract}\noindent
  {
Perturbative calculations at next-to-next-to-leading order
for multi-particle final states require a method to cancel
infrared singularities.
I discuss the subtraction method at NNLO.
As a concrete example I consider the leading-colour contributions
to $e^+ e^- \rightarrow \; \mbox{2 jets}$.
This is the simplest example which exhibits all essential features.
For this example, explicit subtraction terms are given,
which approximate the four-parton and three-parton final states in all
double and single unresolved limits, such that the subtracted matrix
elements can be integrated numerically.
  }
\end{abstract}

\vspace*{\fill}

% footnotes -------------------------------------
 \noindent 
 $^1${\small email address : stefanw@fis.unipr.it}

% main text ------------------------------------
\newpage

%\reversemarginpar

\section{Introduction}

Present and future collider experiments will provide a large sample of multi-particle
final states.
In order to extract information from this data, precise theoretical calculations
are necessary.
This implies to extend perturbative calculations from next-to-leading order (NLO)
to next-to-next-to-leading order (NNLO).
Due to a large variety of interesting jet observables it is desirable not to perform
this calculation for a specific observable, but to set up a computer
program, which yields predictions for any infra-red safe observable relevant to the process
under consideration.
Compared to already existing NNLO calculations for specific (inclusive) observables like
Drell-Yan or inclusive Higgs production
%\cite{Hamberg:1991np,Catani:2001ic,Catani:2001cr,Harlander:2001is,Harlander:2002wh,Anastasiou:2002yz,Ravindran:2003},
\cite{Hamberg:1991np}-\cite{Ravindran:2003},
this requires to work with fully differential cross-sections. This is the major
complication of the project.
The three major challenges to accomplish this task are: the calculation of two-loop amplitudes,
a method for the cancellation of infrared divergences and stable and efficient Monte
Carlo techniques.

The last three years have witnessed a tremendous progress in the calculation of two-loop
amplitudes 
%\cite{Bern:2000ie,Bern:2000dn,Anastasiou:2000kg,Anastasiou:2000ue,Anastasiou:2000mv,Anastasiou:2001sv,Glover:2001af,Bern:2001dg,Bern:2001df,Bern:2002tk}.
\cite{Bern:2000ie}-\cite{Bern:2001dg}.
In particular the two-loop amplitudes for $e^+ e^- \rightarrow \mbox{3 jets}$ have been
calculated
%\cite{Garland:2001tf,Garland:2002ak,Moch:2002hm}.
\cite{Garland:2001tf,Moch:2002hm}.

Up to now, less is known for the cancellation of infrared divergences at NNLO.
Loop amplitudes, calculated in dimensional regularization, 
have explicit poles in the 
dimensional regularization parameter $\eps=2-D/2$,
arising from infrared singularities.
These poles cancel with similar poles arising from
amplitudes with additional partons, when integrated over phase space regions where
two (or more) partons become ``close'' to each other.
However, the cancellation occurs only after the integration over the unresolved phase space
has been performed and prevents thus a naive Monte Carlo approach for a fully exclusive
calculation.
It is therefore necessary to cancel first analytically all infrared divergences and to use
Monte Carlo methods only after this step has been performed.
Infrared divergences occur already at next-to-leading order.
In this case, general methods 
to handle the problem are known.
Examples are
the phase-space slicing method
%\cite{Giele:1992vf,Giele:1993dj,Keller:1998tf}
\cite{Giele:1992vf,Keller:1998tf}
and the subtraction method
%\cite{Frixione:1996ms,Catani:1997vz,Catani:1997vzerr,Dittmaier:1999mb,Phaf:2001gc,Catani:2002hc}.
\cite{Frixione:1996ms}-\cite{Catani:2002hc}.
As already mentionend, similar methods at NNLO are not yet known.
The major difficulty arises from so-called double unresolved
configurations, where three partons become degenerate simultaneously.
A phase-space slicing approach to double unresolved configurations
has been used in the calculation of the photon + 1-jet rate in electron-positron
annihilation \cite{Gehrmann-DeRidder:1998gf}.
The calculational complexity of this process is somewhere in between a NLO calculation and
 a full NNLO calculation, as the relevant two-loop amplitude vanishes.

In this paper I discuss the general setup for the subtraction method at NNLO.
To illustrate the method, I use as an example the leading-colour NNLO
contributions to $e^+ e^- \rightarrow \; \mbox{2 jets}$.
This is the simplest example which exhibits all essential features.
For this example I give all necessary subtraction terms.
These subtraction terms approximate the leading-colour Born contribution
of $e^+ e^- \rightarrow q g g \bar{q}$ in all double and single
unresolved limits, and the leading-colour one-loop contribution
of $e^+ e^- \rightarrow q g \bar{q}$ in all single unresolved
limits.
The subtracted matrix elements can therefore be integrated numerically
over the appropriate phase space.
Valuable information for the construction of the subtraction terms
is provided by the known behaviour of the tree and one-loop
amplitudes in singular limits.
The double unresolved limits of tree amplitudes have been considered in 
%\cite{Berends:1989zn,Catani:1992unpublished,Gehrmann-DeRidder:1998gf,Campbell:1998hg,Catani:1998nv,Catani:1999ss,DelDuca:1999ha,Kosower:2002su}.
\cite{Gehrmann-DeRidder:1998gf}-\cite{Kosower:2002su}.
Single unresolved limits of one-loop amplitudes have been considered in
%\cite{Bern:1994zx,Bern:1998sc,Kosower:1999xi,Kosower:1999rx,Bern:1999ry,Catani:2000pi,Kosower:2003cz}.
\cite{Bern:1994zx}-\cite{Kosower:2003cz}.
However, for the subtraction method this information is not yet sufficient.
(It is sufficient for a phase space slicing approach.) Since the
subtraction terms are subtracted over the complete phase space, they must
have the correct singular behaviour not only in the double unresolved
cases, but also in the single unresolved ones.
This requires to extend the calculations of the singular behaviour
of tree amplitudes and
to include also subleading single unresolved singularities.

The method presented here is general and not restricted to the example
of $e^+ e^- \rightarrow \; \mbox{2 jets}$.
Subtraction terms for other splittings (like $g \rightarrow g g g$)
and other kinematical configuartions (e.g. with partons in the initial state)
can be worked out along the same lines.

The subtraction terms still need to be integrated analytically over the
appropriate unresolved phase space.
Although some of the occuring integrals are quite involved, 
recently developped integration techniques 
%\cite{Gehrmann:2000zt,Gehrmann:2001ck,Moch:2001zr,Weinzierl:2002hv} 
\cite{Gehrmann:2000zt,Moch:2001zr} 
seem to indicate that these integrals can be done analytically.

The knowledge of the subtraction terms will also be useful for 
an improvement of parton showering algorithms in event generators
\cite{Weinzierl:2001ny}.

This paper is organized as follows:
In the following section I discuss the general setup of the subtraction method at NNLO.
Sect. \ref{sectex} gives the relevant amplitudes for the example
$e^+ e^- \rightarrow \; \mbox{2 jets}$.
Sect. \ref{sectlim} reviews the singular limits of amplitudes.
In sect. \ref{sectsubtr} I present the NNLO subtraction terms and
in sect. \ref{sectcanc} I discuss the mechanism of cancellations between
the various terms.
Finally sect. \ref{sectconcl} contains the conclusions and an outlook.

% section ------------------------------------------------------------------

\section{General idea}

\begin{figure}
\begin{center}
\begin{picture}(100,100)(0,0)
\Oval(50,50)(40,10)(30)
\Line(10,10)(29,83.64)
\Line(10,10)(70,15.36)
\Text(27,90)[lb]{$y_{cut}$}
\Line(10,10)(30,30)
\ArrowLine(30,30)(80,80)
\end{picture}
\begin{picture}(100,100)(0,0)
\Oval(50,50)(40,10)(30)
\Line(10,10)(29,83.64)
\Line(10,10)(70,15.36)
\Text(27,90)[lb]{$y_{cut}$}
\Line(10,10)(30,30)
\ArrowLine(30,30)(80,80)
\Gluon(30,30)(90,60){-4}{7}
\end{picture}
\begin{picture}(100,100)(0,0)
\Oval(50,50)(40,10)(30)
\Line(10,10)(29,83.64)
\Line(10,10)(70,15.36)
\Text(27,90)[lb]{$y_{cut}$}
\Line(10,10)(30,30)
\ArrowLine(30,30)(80,80)
\Gluon(20,20)(70,85){4}{7}
\Gluon(30,30)(90,60){-4}{7}
\end{picture}
\end{center}
\caption{\label{jets} Modelling of jets at next-to-next-to-leading order. 
The cone labeled with $y_{cut}$ represents the 
experimental cuts. At next-to-next-to-leading order a jet is modelled either by
one, two or three partons.}
\end{figure}
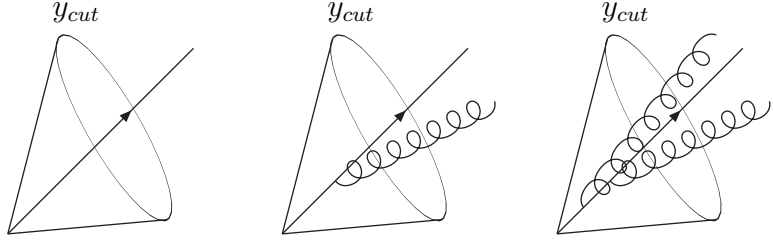

In this section I outline the general setup for the subtraction method at NNLO.
The NNLO contribution to an observable ${\cal O}$ whose LO contribution depends on $n$ partons
is given by
\bq
\lefteqn{
\l {\cal O} \r_n^{NNLO} = } \\ 
& &
   \int {\cal O}_{n+2}(p_1,...,p_{n+2}) \; d\sigma_{n+2}^{(0)} 
 + \int {\cal O}_{n+1}(p_1,...,p_{n+1}) \; d\sigma_{n+1}^{(1)} 
 + \int {\cal O}_{n}(p_1,...,p_{n}) \; d\sigma_n^{(2)}. \nonumber 
\eq
The various contributions are symbolically shown in fig. \ref{jets}.
The observable ${\cal O}$ has to be infrared safe. This requires that whenever a 
$n+1$ parton configuration $p_1$,...,$p_{n+1}$ becomes kinematically degenerate 
with a $n$ parton configuration $p_1'$,...,$p_{n}'$
we must have
\bq
{\cal O}_{n+1}(p_1,...,p_{n+1}) & \rightarrow & {\cal O}_n(p_1',...,p_n').
\eq
In addition, we must have in the double unresolved case (e.g. when
a $n+2$ parton configuration $p_1$,...,$p_{n+2}$ becomes kinematically degenerate 
with a $n$ parton configuration $p_1'$,...,$p_{n}'$)
\bq
{\cal O}_{n+2}(p_1,...,p_{n+2}) & \rightarrow & {\cal O}_n(p_1',...,p_n').
\eq
The subscript $n$ for ${\cal O}_{n}$ and $d\sigma_n^{(l)}$
indicates that this contribution is integrated over a $n$-parton phase space.
The various contributions $d\sigma_n^{(l)}$ are given by
\bq
d\sigma_{n+2}^{(0)} & = & 
 \left( \left. {\cal A}_{n+2}^{(0)} \right.^\ast {\cal A}_{n+2}^{(0)} \right) 
d\phi_{n+2},  \nonumber \\
d\sigma_{n+1}^{(1)} & = & 
 \left( 
 \left. {\cal A}_{n+1}^{(0)} \right.^\ast {\cal A}_{n+1}^{(1)} 
 + \left. {\cal A}_{n+1}^{(1)} \right.^\ast {\cal A}_{n+1}^{(0)} \right)  
d\phi_{n+1}, \nonumber \\
d\sigma_n^{(2)} & = & 
 \left( 
 \left. {\cal A}_n^{(0)} \right.^\ast {\cal A}_n^{(2)} 
 + \left. {\cal A}_n^{(2)} \right.^\ast {\cal A}_n^{(0)}  
 + \left. {\cal A}_n^{(1)} \right.^\ast {\cal A}_n^{(1)} \right) d\phi_n,
\eq
where ${\cal A}_n^{(l)}$ denotes an amplitude with $n$ external partons and $l$ loops.
$d\phi_n$ is the phase space measure for $n$ partons.
Taken separately, each of the individual contributions 
$d\sigma_{n+2}^{(0)}$, $d\sigma_{n+1}^{(1)}$ and $d\sigma_{n}^{(2)}$  
gives a divergent contribution.
Only the sum of all contributions is finite.
However, since one would like to perform the integration numerically, one must
first cancel analytically the divergences, before numerical integration over
the different phase spaces can be envisaged.
This is not a new phenomena at NNLO, it already occurs at NLO.
We can consider an observable, whose LO contribution depends on $n+1$ partons.
The NLO contribution is given by
\bq
\l {\cal O} \r_{n+1}^{NLO} 
& = &
   \int {\cal O}_{n+2}(p_1,...,p_{n+2}) \; d\sigma_{n+2}^{(0)} 
 + \int {\cal O}_{n+1}(p_1,...,p_{n+1}) \; d\sigma_{n+1}^{(1)}. 
\eq
To render the two contributions finite, one adds and subtracts a suitable term
\cite{Catani:1997vz}
\bq
\lefteqn{
\int {\cal O}_{n+2} \; d\sigma_{n+2}^{(0)} + \int {\cal O}_{n+2} \; d\sigma_{n+1}^{(1)} = } \nonumber \\ 
& &
\int \left( {\cal O}_{n+2} \; d\sigma_{n+2}^{(0)} - {\cal O}_{n+1} \circ d\alpha^{(0,1)}_{n+1} \right) 
+ \int \left( {\cal O}_{n+1} \; d\sigma_{n+1}^{(1)} + {\cal O}_{n+1} \circ d\alpha^{(0,1)}_{n+1} \right).
\eq
The approximation $d\alpha^{(0,1)}_{n+1}$ has to fulfill the following requirements:
First, $d\alpha^{(0,1)}_{n+1}$ must be a proper approximation of $d\sigma_{n+2}^{(0)}$ such as to have the same pointwise singular behaviour
in $D$ dimensions as $d\sigma_{n+2}^{(0)}$ itself. Thus, $d\alpha^{(0,1)}_{n+1}$ acts as a local counterterm for $d\sigma_{n+2}^{(0)}$ and one
can safely perform the limit $\varepsilon \rightarrow 0$. 
Secondly, $d\sigma_{n+1}^{(0,1)}$ must be analytically integrable in $D$ dimensions over the one-parton subspace leading to soft and 
collinear divergences. 
The integral
\bq
\label{onepartonsubspace}
\int_1 d\alpha_{n+1}^{(0,1)} 
\eq
contains then explicit poles in $1/\eps^2$ and $1/\eps$ which exactly cancel the poles
from the loop amplitude in $d\sigma_{n+1}^{(1)}$.
Note that the observable ${\cal O}$ in the approximation term is evaluated with a $n+1$
parton configuration. This allows the integration over the one-parton subspace to be
performed analytically, but requires a mapping of four-momenta from the $n+2$ parton
configuration to the $n+1$ parton configuration.
This mapping has to satisfy momentum conservation and also has to keep all partons massless.
In addition it has to have the right behaviour in the singular limits.
Several choices for such a mapping exist \cite{Catani:1997vz,Kosower:1998zr}.
The notation ${\cal O}_{n+1} \circ d\alpha^{(0,1)}_{n+1}$ is a reminder, that
in general the approximation is a sum of terms
\bq
{\cal O}_{n+1} \circ d\alpha^{(0,1)}_{n+1} & = & \sum {\cal O}_{n+1} \; d\alpha^{(0,1)}_{n+1}
\eq
and the mapping used to relate the $n+2$ parton configuration to a $n+1$ parton configuration
differs in general for each summand.
\\
\\
In a similar way, I rewrite the NNLO contribution as
\bq
\l {\cal O} \r_n^{NNLO} & = &
 \int \left( {\cal O}_{n+2} \; d\sigma_{n+2}^{(0)} 
             - {\cal O}_{n+1} \circ d\alpha^{(0,1)}_{n+1}
             - {\cal O}_{n} \circ d\alpha^{(0,2)}_{n} 
      \right) \nonumber \\
& &
 + \int \left( {\cal O}_{n+1} \; d\sigma_{n+1}^{(1)} 
               + {\cal O}_{n+1} \circ d\alpha^{(0,1)}_{n+1}
               - {\cal O}_{n} \circ d\alpha^{(1,1)}_{n}
        \right) \nonumber \\
& & 
 + \int \left( {\cal O}_{n} \; d\sigma_n^{(2)} 
               + {\cal O}_{n} \circ d\alpha^{(0,2)}_{n}
               + {\cal O}_{n} \circ d\alpha^{(1,1)}_{n}
        \right).
\eq
Here $d\alpha_n^{(0,2)}$ acts as a local counterterm 
for double unresolved configurations 
to $d\sigma_{n+2}^{(0)} - d\alpha^{(0,1)}_{n+1}$ in $D$
dimensions.
Furthermore, it is integrable over a two-parton subspace.
$d\alpha_n^{(1,1)}$ is a local counterterm 
to $d\sigma_{n+1}^{(1)} + d\alpha^{(0,1)}_{n+1}$ 
in $D$ dimensions,
whenever one of the $(n+1)$ partons becomes unresolved.
It is integrable over a one-parton subspace.
This setup extends to higher orders and in general,
$d\alpha^{(l,k)}_{n}$ is an approximation to 
\bq
d\sigma_{n+k}^{(l)}
+ d\alpha_{n+k}^{(0,l)} + d\alpha_{n+k}^{(1,l-1)} + ... + d\alpha_{n+k}^{(l-1,1)}
- d\alpha_{n+k-1}^{(l,1)} - d\alpha_{n+k-2}^{(l,2)} - ... - d\alpha_{n+1}^{(l,k-1)},
\eq
e.g. a contribution with 
at most $n+k$ partons and at most $l$ loops in the case when $k$ partons become unresolved.
\\
The task at NNLO is to give the explicit forms of $d\alpha_{n}^{(0,2)}$ 
(and of $d\alpha_{n}^{(0,1)}$ and $d\alpha_{n}^{(1,1)}$) and to integrate these approximations over
a two-parton phase space (or a one-parton phase space in the case of
$d\alpha_{n}^{(0,1)}$ and $d\alpha_{n}^{(1,1)}$).
\\
\\
To illustrate the cancellations for double unresolved contributions, I consider a simple
toy model, where the double real emission contribution 
to a 1-jet observable is given by
\bq
{\cal O}_{3} \; d\sigma_{3}^{(0)} & = & \theta_3(x_1,x_2,x_3) F_3(x_1,x_2,x_3).
\eq
The singular behaviour of the functions $F_3$ is assumed to be
\bq
\lim\limits_{x_1\rightarrow 0} F_3(x_1,x_2,x_3) & = & \frac{1}{x_1} F_2(x_2,x_3), \nonumber \\
\lim\limits_{x_1\rightarrow 0} F_2(x_1,x_2) & = & \frac{1}{x_1} F_1(x_2).
\eq
The function $\theta_3$, defining the observable satisfies
\bq
\lim\limits_{x_1\rightarrow 0} \theta_3(x_1,x_2,x_3) & = & \theta_2(x_2,x_3), \nonumber \\
\lim\limits_{x_1\rightarrow 0} \theta_2(x_1,x_2) & = & \theta_1(x_2).
\eq
Similar relations are assumed to hold for the singular limits in the other variables.
The NLO subtraction term is given by
\bq
\lefteqn{
{\cal O}_{2} \circ d\alpha_{2}^{(0,1)} = } \nonumber \\
& & 
 \frac{1}{x_1} F_2(x_2,x_3) \theta_2(x_2,x_3)
 + \frac{1}{x_2} F_2(x_1,x_3) \theta_2(x_1,x_3)
 + \frac{1}{x_3} F_2(x_1,x_2) \theta_2(x_1,x_2).
\eq
Let us first consider the cancellation of single unresolved
contributions to a 2-jet observable. I consider the $x_1 \rightarrow 0$ limit.
Here, ${\cal O}_{3} \; d\sigma_{3}^{(0)} - {\cal O}_{2} \circ d\alpha_{2}^{(0,1)}$ consists of two contributions: The first part
\bq
\lim\limits_{x_1\rightarrow 0}
\left(
F_3(x_1,x_2,x_3) \theta_3(x_1,x_2,x_3) - \frac{1}{x_1} F_2(x_2,x_3) \theta_2(x_2,x_3) \right)
\eq
is finite by construction.
The second part is given by
\bq
\label{divsingleunres}
\lefteqn{
\hspace*{-3cm}
\lim\limits_{x_1\rightarrow 0}
 \left(
 - \frac{1}{x_2} F_2(x_1,x_3) \theta_2(x_1,x_3)
 - \frac{1}{x_3} F_2(x_1,x_2) \theta_2(x_1,x_2)
 \right) = } \nonumber \\
 & &
 - \frac{1}{x_1 x_2} F_1(x_3) \theta_1(x_3)
 - \frac{1}{x_1 x_3} F_1(x_2) \theta_1(x_2).
\eq
This part is proportional to $\theta_1(x_2)$ or $\theta_1(x_3)$ and 
therefore does not contribute to a 2-jet observable ($\theta_1$ is zero
for a two-jet observable).
However, this argument no longer holds if we move to NNLO and consider
double unresolved contributions to a 1-jet observable.
Here, $\theta_1$ does not vanish and the r.h.s of eq. (\ref{divsingleunres})
gives a divergent contribution already in the single unresolved case
$x_1\rightarrow 0$, and $x_2, x_3 \neq 0$.
At NNLO these singular pieces have to be taken properly into account and subtracted out.
To summarize, the major complication for the subtraction method is not to find
subtraction terms which match the matrix element in all double unresolved limits,
but to find subtraction terms, which match in all double unresolved limits
{\bf and} all single unresolved limits.
For the toy example the NNLO subtraction term is given by
\bq
\lefteqn{
{\cal O}_{1} \circ d\alpha_{2}^{(0,2)} = } \nonumber \\
& & 
 - \frac{1}{x_1 x_2} F_1(x_3) \theta_1(x_3)
 - \frac{1}{x_1 x_3} F_1(x_2) \theta_1(x_2)
 - \frac{1}{x_2 x_3} F_1(x_1) \theta_1(x_1).
\eq
Note that in this toy example the NNLO subtraction term can be obtained
solely from the singular behaviour of the NLO subtraction term.
In general however, the limits do not commute, e.g. taking the singular limit
of the single unresolved approximation does not equal the double unresolved approximation
of the matrix element.

% section ------------------------------------------------------------------

\section{An example}
\label{sectex}

To make the method for the double unresolved contributions 
as transparent as possible, I consider as a simple
example the leading $N_c$-contribution 
to the tree-level amplitude $e^+ e^- \rightarrow q g g \bar{q}$.
This example is complicated enough to expose all essential features.

\subsection{Double real emission}

The singular behaviour of the kinematical structure of QCD amplitudes 
is entangled with colour factors.
For a clear understanding of the factorization properties in soft and collinear limits
it is desirable to disentangle the colour factors from the rest.
This can be done either by introducing colour charge operators \cite{Catani:1997vz}
or by decomposing the full amplitude into partial and primitive amplitudes.
In this paper I follow the second approach, since this method is closer related
to the way several one-loop amplitudes were calculated.
Throughout this paper I use the following normalization for the colour matrices:
\bq
\mbox{Tr} \; T^a T^b = \frac{1}{2} \delta^{ab}.
\eq
Tree-level amplitudes are decomposed into colour factors and partial amplitudes.
Each partial amplitude is gauge-invariant and has usually a simpler structure.
For example, the colour decompostion of the the tree amplitudes for
$e^+ e^- \rightarrow q_1 g_2 g_3 \bar{q}_4$ amplitude reads
\bq
\lefteqn{
{\cal A}^{(0)}_4(q_1,g_2,g_3,\bar{q}_4) = } \nonumber \\
& & e^2 g^2 c_0
 \left[ \left( T^2 T^3 \right)_{14} A^{(0)}_4(q_1,g_2,g_3,\bar{q}_4)
       +
        \left( T^3 T^2 \right)_{14} A^{(0)}_4(q_1,g_3,g_2,\bar{q}_4)
 \right].
\eq
Here,
$c_0$ denotes a factor from the electro-magnetic coupling:
\begin{eqnarray}
c_0 & = & -Q^q + v^e v^q {\cal P}_Z(s), \nonumber \\
{\cal P}_Z(s) & = & \frac{s}{s-M_Z^2+ i \Gamma_Z M_Z}.
\end{eqnarray}
The electron -- positron pair can either annihilate into a photon or a $Z$-boson. The first term in the expression
for $c_0$ corresponds to an intermediate photon, whereas the last term corresponds to a $Z$-boson.
The left- and right-handed couplings of the $Z$-boson to fermions are given by
\begin{eqnarray}
v_L = \frac{I_3 - Q \sin^2 \theta_W}{\sin \theta_W \cos \theta_W}, & &
v_R = \frac{- Q \sin \theta_W}{\cos \theta_W},
\end{eqnarray}
where $Q$ and $I_3$ are the charge and the third component of the weak isospin of the fermion. 
The short-hand notation for the colour-matrices corresponds to 
$T^2_{14} = T^{a_g}_{i_q j_{\bar{q}}}$.
\begin{figure}
\begin{center}
\begin{tabular}{ccc}
\begin{picture}(110,100)(0,0)
 \ArrowLine(20,50)(0,70)
 \ArrowLine(0,30)(20,50)
% \ArrowLine(0,70)(20,50)
% \ArrowLine(20,50)(0,30)
 \Vertex(20,50){2}
 \Photon(20,50)(50,50){4}{3}
 \Vertex(50,50){2}
 \ArrowLine(90,10)(50,50)
 \ArrowLine(50,50)(90,90)
 \Vertex(60,60){2}
 \Vertex(75,75){2}
 \Gluon(60,60)(90,60){-4}{3}
 \Gluon(75,75)(90,75){-4}{1}
 \Text(50,0)[b]{Diagram 1a}
 \Text(95,90)[l]{$p_1$}
 \Text(95,75)[l]{$p_2$}
 \Text(95,60)[l]{$p_3$}
 \Text(95,10)[l]{$p_4$}
\end{picture}
&
\begin{picture}(110,100)(0,0)
 \ArrowLine(20,50)(0,70)
 \ArrowLine(0,30)(20,50)
% \ArrowLine(0,70)(20,50)
% \ArrowLine(20,50)(0,30)
 \Vertex(20,50){2}
 \Photon(20,50)(50,50){4}{3}
 \Vertex(50,50){2}
 \ArrowLine(90,10)(50,50)
 \ArrowLine(50,50)(90,90)
 \Vertex(75,25){2}
 \Vertex(75,75){2}
 \Gluon(75,25)(90,25){4}{1}
 \Gluon(75,75)(90,75){-4}{1}
 \Text(50,0)[b]{Diagram 1b}
 \Text(95,90)[l]{$p_1$}
 \Text(95,75)[l]{$p_2$}
 \Text(95,25)[l]{$p_3$}
 \Text(95,10)[l]{$p_4$}
\end{picture}
&
\begin{picture}(110,100)(0,0)
 \ArrowLine(20,50)(0,70)
 \ArrowLine(0,30)(20,50)
% \ArrowLine(0,70)(20,50)
% \ArrowLine(20,50)(0,30)
 \Vertex(20,50){2}
 \Photon(20,50)(50,50){4}{3}
 \Vertex(50,50){2}
 \ArrowLine(90,10)(50,50)
 \ArrowLine(50,50)(90,90)
 \Vertex(75,25){2}
 \Vertex(60,40){2}
 \Gluon(75,25)(90,25){4}{1}
 \Gluon(60,40)(90,40){4}{3}
 \Text(50,0)[b]{Diagram 1c}
 \Text(95,90)[l]{$p_1$}
 \Text(95,40)[l]{$p_2$}
 \Text(95,25)[l]{$p_3$}
 \Text(95,10)[l]{$p_4$}
\end{picture}
\\
\end{tabular}
\begin{tabular}{cc}
\begin{picture}(110,100)(0,0)
 \ArrowLine(20,50)(0,70)
 \ArrowLine(0,30)(20,50)
% \ArrowLine(0,70)(20,50)
% \ArrowLine(20,50)(0,30)
 \Vertex(20,50){2}
 \Photon(20,50)(50,50){4}{3}
 \Vertex(50,50){2}
 \ArrowLine(90,10)(50,50)
 \ArrowLine(50,50)(90,90)
 \Vertex(60,60){2}
 \Gluon(60,60)(75,60){-4}{1}
 \Vertex(75,60){2}
 \Gluon(75,60)(90,75){-4}{2}
 \Gluon(75,60)(90,45){-4}{2}
 \Text(50,0)[b]{Diagram 1d}
 \Text(95,90)[l]{$p_1$}
 \Text(95,75)[l]{$p_2$}
 \Text(95,45)[l]{$p_3$}
 \Text(95,10)[l]{$p_4$}
\end{picture}
&
\begin{picture}(110,100)(0,0)
 \ArrowLine(20,50)(0,70)
 \ArrowLine(0,30)(20,50)
% \ArrowLine(0,70)(20,50)
% \ArrowLine(20,50)(0,30)
 \Vertex(20,50){2}
 \Photon(20,50)(50,50){4}{3}
 \Vertex(50,50){2}
 \ArrowLine(90,10)(50,50)
 \ArrowLine(50,50)(90,90)
 \Vertex(60,40){2}
 \Gluon(60,40)(75,40){4}{1}
 \Vertex(75,40){2}
 \Gluon(75,40)(90,55){4}{2}
 \Gluon(75,40)(90,25){4}{2}
 \Text(50,0)[b]{Diagram 1e}
 \Text(95,90)[l]{$p_1$}
 \Text(95,55)[l]{$p_2$}
 \Text(95,25)[l]{$p_3$}
 \Text(95,10)[l]{$p_4$}
\end{picture}
\\
\end{tabular}
\caption{\label{diagrams} Diagrams contributing to the colour-ordered partial
amplitude $A^{(0)}_4(q_1,g_2,g_3,\bar{q}_4)$.}
\end{center}
\end{figure}
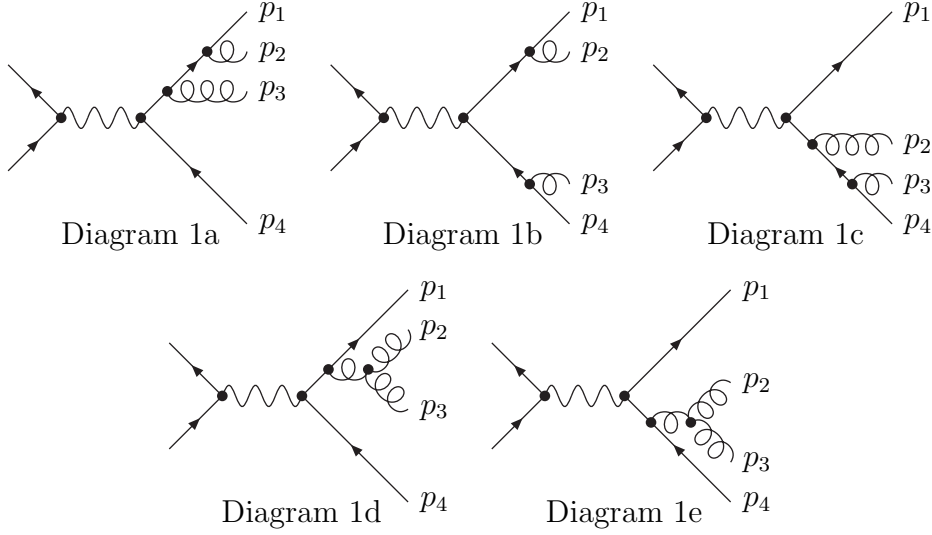
Figure (\ref{diagrams}) shows the diagrams contributing to the colour-ordered 
partial amplitude $A^{(0)}_4(q_1,g_2,g_3,\bar{q}_4)$.
Squaring the full amplitude ${\cal A}^{(0)}_4(q_1,g_2,g_3,\bar{q}_4)$ one obtains
\bq
\label{squared}
\lefteqn{
\left. {\cal A}^{(0)}_4 \right.^\ast {\cal A}^{(0)}_4 =  
 \left( 4 \pi \alpha \right)^2 \left( 4 \pi \alpha_s \right)^2 
 \left| c_0 \right|^2
} \nonumber \\ 
& &
 \times 
 \frac{1}{4} \left( N_c^2 -1 \right) N_c 
 \left[
   \left| A^{(0)}_4(q_1,g_2,g_3,\bar{q}_4) \right|^2
   + \left| A^{(0)}_4(q_1,g_3,g_2,\bar{q}_4) \right|^2
 + O \left( \frac{1}{N_c^2} \right) 
 \right].
\eq
Note that interference terms between $A^{(0)}_4(q_1,g_2,g_3,\bar{q}_4)$ and
$A^{(0)}_4(q_1,g_3,g_2,\bar{q}_4)$ are subleading in $N_c$.
In four dimensions compact expressions 
for the partial amplitudes are obtained by calculating
helicity amplitudes.
For the partial amplitude $A^{(0)}_4(q_1,g_2,g_3,\bar{q}_4)$
there are only three independent helicity
configurations, which can be taken to be
\bq
q_1^+,g_2^+,g_3^+,\bar{q}_4^-,p_5^-,p_6^+, \;\;\;
q_1^+,g_2^+,g_3^-,\bar{q}_4^-,p_5^-,p_6^+, \;\;\;
q_1^+,g_2^-,g_3^+,\bar{q}_4^-,p_5^-,p_6^+.
\eq
The helicity amplitudes are expressed in term of spinorproducts
and read \cite{Bern:1997sc}:
\bq
\lefteqn{
A^{(0)}_4(q_1^+,g_2^+,g_3^+,\bar{q}_4^-,p_5^-,p_6^+) = 
- 4 \,i\, \frac{\spa4.5^2}{\spa1.2\spa2.3\spa3.4\spa5.6}, }
\\
\lefteqn{
A^{(0)}_4(q_1^+,g_2^+,g_3^-,\bar{q}_4^-,p_5^-,p_6^+) = 
4 \, i \left[  
   \frac{\spa3.1 \spb1.2 \spa4.5 \spab3.{(1+2)}.6 
       }{ \spa1.2 s_{23} s_{123} s_{56} }
 \right.
   } \nonumber \\
& & 
 \hspace*{35mm}
 \left.
  - \frac{\spa3.4\spb4.2 \spb1.6 \spab5.{(3+4)}.2
       }{ \spb3.4 s_{23} s_{234} s_{56} } 
  - \frac{ \spab5.{(3+4)}.2 \, \spab3.{(1+2)}.6
       }{ \spa1.2 \spb3.4 s_{23} s_{56} } \right], 
\nonumber \\
\lefteqn{
A^{(0)}_4(q_1^+,g_2^-,g_3^+,\bar{q}_4^-,p_5^-,p_6^+) = } \nonumber \\
& &
4 \, i \, \left[
- \frac{\spb1.3^2\spa4.5 \spab2.{(1+3)}.6 }{ \spb1.2 s_{23} s_{123} s_{56}}
+ \frac{\spa2.4^2\spb1.6 \spab5.{(2+4)}.3 }{ \spa3.4 s_{23} s_{234} s_{56}}
+ \frac{\spb1.3\spa2.4\spb1.6\spa4.5 }{ \spb1.2\spa3.4 s_{23} s_{56}}
   \right].
\nonumber 
\eq
The notation of spinor products follows \cite{Bern:1997sc}.
For the singular limits it is sufficient to know that spinor products
scale as the square root of invariants $s_{ij}=(p_i+p_j)^2$:
\bq
\left| \spa{i}.{j} \right| = \left| \spb{i}.{j} \right| 
 = \sqrt{s_{ij}}
\eq
For the last two helicity amplitudes there are alternate forms, which have
a clearer singular structure in the $p_2 || p_3$ limit, at the expense
of a more obscure behaviour in the triple collinear limits
$p_1 || p_2 || p_3$ and $p_2 || p_3 || p_4$: 
\bq
\lefteqn{
A^{(0)}_4(q_1^+,g_2^+,g_3^-,\bar{q}_4^-,p_5^-,p_6^+) = 
4 \, i \left[
   \frac{ \spa5.4 \spb4.2 \spb1.2 \, \spab5.{(3+4)}.2
      }{ \spb2.3\spb3.4 s_{123}s_{234} \spa5.6 }
 \right.
} \\
& & 
 \hspace*{30mm}
 \left.
 + \frac{ \spa3.1\spb1.6 \spa3.4 \, \spab3.{(1+2)}.6
      }{ \spa1.2\spa2.3 s_{123}s_{234} \spb5.6 } 
 - \frac{ \spab3.{(1+2)}.6 \, \spab5.{(3+4)}.2
      }{ \spa1.2\spb3.4 s_{123}s_{234} } \right], 
 \nonumber \\
\lefteqn{
A^{(0)}_4(q_1^+,g_2^-,g_3^+,\bar{q}_4^-,p_5^-,p_6^+) = } \nonumber \\
& &
 4 \, i \left[
   \frac{ \spb1.3^2 \spa4.5 \spab5.{(2+4)}.3
       }{ \spb1.2 \spb2.3 s_{123} s_{234} \spa5.6 }
  - \frac{ \spa2.4^2 \spb1.6 \spab2.{(1+3)}.6
       }{ \spa2.3 \spa3.4 s_{123} s_{234} \spb5.6 }
  + \frac{ \spb1.3 \spa2.4 \spb1.6 \spa4.5 
       }{ \spb1.2 \spa3.4 s_{123} s_{234} } \right].
 \nonumber
\eq
As can be seen from these explicit formulae there
are simultaneous singularities in single collinear limits
$s_{ij} \rightarrow 0$ and triple collinear limits
$s_{ijk} \rightarrow 0$.
These are nested singularities.
It should be noted that the helicity amplitudes are sufficient
for the derivation of the subtraction terms when working
in a four-dimensional regularization scheme 
%\cite{Bern:1992aq,Weinzierl:1999xb,Weinzierl:2000fe,Bern:2002zk}.
\cite{Bern:1992aq}-\cite{Bern:2002zk}.
In conventional dimensional regularization \cite{Collins}
the squared matrix elements
has $O(\eps)$-terms, which also have to be taken into account.

\subsection{One-loop amplitudes with one unresolved parton}

For a NNLO analysis of $e^+ e^- \rightarrow \;\mbox{2 jets}$ 
one needs also the amplitude for $e^+ e^- \rightarrow q g \bar{q}$ up to 
one loop.
The colour decomposition of the tree amplitudes and the one-loop amplitudes are
\bq
{\cal A}^{(0)}_3(q_1,g_2,\bar{q}_3)  & = &
 e^2 g c_0 
 \left( T^2 \right)_{13} A^{(0)}_3(q_1,g_2,\bar{q}_3),
 \nonumber \\
{\cal A}^{(1)}_3(q_1,g_2,\bar{q}_3)  & = &
 e^2 g c_0 
        \left( T^2 \right)_{13} \left( \frac{\alpha_s}{2\pi} \right)
        A^{(1)}_{3,\mbox{\tiny partial}}(q_1,g_2,\bar{q}_3).
\eq
For both the tree amplitude and the one-loop amplitude there is only one partial
amplitude.
However, to obtain a clear factorization in singular kinematical limits, it is convenient
to decompose the one-loop partial amplitude further into primitive amplitudes
\cite{Bern:1999ry}.
Primitive amplitudes can be defined as the sum of all Feynman diagrams with a fixed cyclic ordering
of the QCD partons and a definite routing of the external fermion lines through the diagram.
For the case at hand we have
\bq
\lefteqn{
A^{(1)}_{3,\mbox{\tiny partial}}(q_1,g_2,\bar{q}_3) = } \nonumber \\
 & &
 \frac{1}{2} N_c A^{(1),L,[1]}_3(q_1,g_2,\bar{q}_3) 
                              - \frac{1}{2 N_c} A^{(1),R,[1]}_3(q_1,g_2,\bar{q}_3)
                              + \frac{1}{2} N_f A^{(1),L,[1/2]}_3(q_1,g_2,\bar{q}_3).
\eq
The superscript $L$ or $R$ indicates whether the external fermion line turns left or right
when entering the loop, while the superscripts $[1]$ and $[1/2]$ indicate the spin of the particle
circulating in the loop.
\begin{figure}
\begin{center}
\begin{tabular}{ccc}
\begin{picture}(135,100)(0,0)
 \ArrowLine(20,50)(0,70)
 \ArrowLine(0,30)(20,50)
% \ArrowLine(0,70)(20,50)
% \ArrowLine(20,50)(0,30)
 \Vertex(20,50){2}
 \Photon(20,50)(50,50){4}{3}
 \Vertex(50,50){2}
 \ArrowLine(90,10)(50,50)
 \ArrowLine(50,50)(90,90)
 \Vertex(75,25){2}
 \Vertex(75,75){2}
 \Gluon(75,75)(100,50){-4}{3}
 \Gluon(75,25)(100,50){4}{3}
 \Vertex(100,50){2}
 \Gluon(100,50)(115,50){-4}{1}
 \Text(50,0)[b]{Diagram 2a}
 \Text(95,90)[l]{$p_1$}
 \Text(120,50)[l]{$p_2$}
 \Text(95,10)[l]{$p_3$}
\end{picture}
&
\begin{picture}(110,100)(0,0)
 \ArrowLine(20,50)(0,70)
 \ArrowLine(0,30)(20,50)
% \ArrowLine(0,70)(20,50)
% \ArrowLine(20,50)(0,30)
 \Vertex(20,50){2}
 \Photon(20,50)(50,50){4}{3}
 \Vertex(50,50){2}
 \ArrowLine(90,10)(50,50)
 \ArrowLine(50,50)(90,90)
 \Vertex(65,65){2}
 \Gluon(65,65)(90,65){-4}{3}
 \Vertex(80.6,80.6){2}
 \Vertex(59.4,59.4){2}
 \GlueArc(70,70)(15,45,225){4}{5}
 \Text(50,0)[b]{Diagram 2b}
 \Text(95,90)[l]{$p_1$}
 \Text(95,65)[l]{$p_2$}
 \Text(95,10)[l]{$p_3$}
\end{picture}
&
\begin{picture}(135,100)(0,0)
 \ArrowLine(20,50)(0,70)
 \ArrowLine(0,30)(20,50)
% \ArrowLine(0,70)(20,50)
% \ArrowLine(20,50)(0,30)
 \Vertex(20,50){2}
 \Photon(20,50)(50,50){4}{3}
 \Vertex(50,50){2}
 \ArrowLine(90,10)(50,50)
 \ArrowLine(50,50)(90,90)
 \Vertex(55,55){2}
 \Gluon(55,55)(78,55){-4}{2}
 \Gluon(92,55)(115,55){-4}{2}
 \Vertex(78,55){2}
 \Vertex(92,55){2}
 \ArrowArc(85,55)(7,0,180)
 \ArrowArc(85,55)(7,180,0)
 \Text(50,0)[b]{Diagram 2c}
 \Text(95,90)[l]{$p_1$}
 \Text(120,40)[l]{$p_2$}
 \Text(95,10)[l]{$p_3$}
\end{picture}
\\
\end{tabular}
\caption{\label{diagramsoneloop} Examples of one-loop diagrams, contributing to different
primitive amplitudes: 
Diagram 2a contributes to $A_3^{(1),L,[1]}$,
diagram 2b contributes to $A_3^{(1),R,[1]}$ and
diagram 2c contributes to $A_3^{(1),L,[1/2]}$.}
\end{center}
\end{figure}
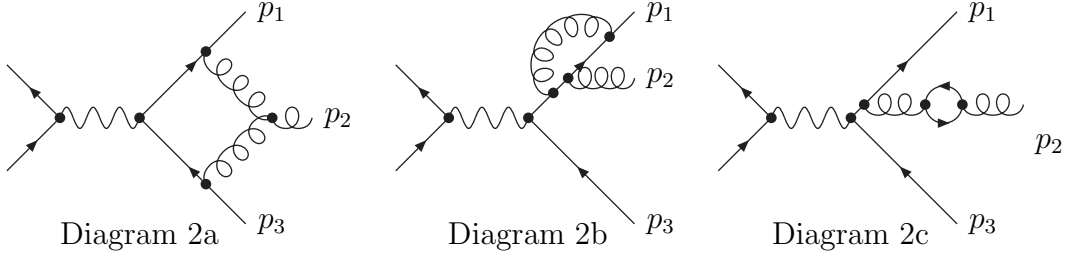
Examples of one-loop diagrams corresponding to different primitive amplitudes are shown in 
fig. (\ref{diagramsoneloop}).
Since we are here only interested in the leading-colour contribution, only the primitive
amplitude $A^{(1),L,[1]}_3(q_1,g_2,\bar{q}_3)$ is relevant.
In later sections I will drop the superscript ``$L,[1]$'' for the leading-colour
primitive amplitude.
There is only one independent helicity configuration, which we can
take to be
\bq
q_1^+,g_2^+,\bar{q}_3^-,p_4^-,p_5^+.
\eq
The helicity amplitudes are expressed in terms of spinorproducts
and read:
% This fixes the sign of the strong coupling g:
% The Feynman rule for the colour-ordered quark-gluon vertex is :
%
%   -i g gamma_mu
%
% and the colour-ordered three-gluon vertex 
% is therefore (momenta outgoing)
%
%    i (   g_(nu,lambda) * ( k3(mu) - k2(mu) ) 
%        + g_(lambda,mu) * ( k1(nu) - k3(nu) ) 
%        + g_(mu,nu) * ( k2(lambda) - k1(lambda) ) 
%      )
%
% This convention has been used in hep-ph/0207043 and hep-ph/9708239
\bq
A^{(0)}_3(q_1^+,g_2^+,\bar{q}_3^-,p_4^-,p_5^+) & = &  
 2 \sqrt{2} i \frac{\l 34 \r^2}{\l 12 \r \l 23 \r \l 54 \r}, 
 \nonumber \\
A^{(1),L,[1]}_3(q_1^+,g_2^+,\bar{q}_3^-,p_4^-,p_5^+) 
 & = & 
 V^{(1)}_3 A^{(0)}_3(q_1^+,g_2^+,\bar{q}_3^-,p_4^-,p_5^+)
 \nonumber \\
 & &
  + 2 \sqrt{2} i \frac{\l 13 \r [ 15 ] \l 34 \r }{\l 12 \r \l 23 \r s_{45}} F_0
  + 2 \sqrt{2} i \frac{\l 13 \r^2 [ 15 ]^2 \l 45 \r}
                      {\l 12 \r \l 23 \r s_{45}^2} F_1.
\eq
The renormalized coefficients $V^{(1)}_3$, $F_0$, and $F_1$ are given in
the HV scheme by
\bq
V^{(1)}_3 & = &
            -\frac{2}{\eps^2}
            + \left( \ln(x_1) + \ln(x_2) - \frac{10}{3} \right) 
                   \frac{1}{\eps}
            - R(x_1,x_2) 
            - \frac{1}{2} \ln^2(x_1)
            - \frac{1}{2} \ln^2(x_2)
            + \frac{7}{6} \pi^2
 \nonumber \\
 & &
            + \frac{3}{2} \ln(x_2)
            - \frac{7}{2}
             + O(\eps) + \mbox{imaginary parts},
 \nonumber \\
F_0 & = &
           \frac{\ln(x_2)}{\left( 1-x_2 \right)} 
           + O(\eps),
 \nonumber \\
F_1 & = &
           \frac{1}{2} \frac{(1-x_2)+\ln(x_2)}{\left( 1-x_2 \right)^2}
           + O(\eps).
\eq
Here, $x_1 = s_{12}/s_{123}$, $x_2 = s_{23}/s_{123}$ and 
the symmetric function $R(x_1,x_2)$ is defined by
\bq
R(x_1,x_2) = \left( \frac{1}{2} \ln(x_1) \ln(x_2)
                -\ln(x_1) \ln(1-x_1)
                +\frac{1}{2} \zeta_2-\mbox{Li}_{2}(x_1) \right) + (x_1 \leftrightarrow x_2)\, .
\eq
The amplitude in the FDH scheme is obtained by replacing $V^{(1)}_3$ by
\bq
V^{(1)}_{3,FDH} & = & 
 V^{(1)}_3 + \frac{1}{2}.
\eq

\subsection{The two-loop amplitude}

For a NNLO calculation one also needs the amplitudes
for $e^+ e^- \rightarrow q \bar{q}$ up to two loops.
The colour decompositions of the tree amplitudes, the one-loop amplitudes and the 
two-loop amplitudes are
\bq
{\cal A}^{(0)}_2(q_1,\bar{q}_2)  & = &
 e^2 c_0 
 \delta_{12} A^{(0)}_2(q_1,\bar{q}_2), 
 \nonumber \\
{\cal A}^{(1)}_2(q_1,\bar{q}_2)  & = &
 e^2 c_0 
 \delta_{12} \left( \frac{\alpha_s}{2\pi} \right) A^{(1)}_{2,\mbox{\tiny partial}}(q_1,\bar{q}_2), 
 \nonumber \\
{\cal A}^{(2)}_2(q_1,\bar{q}_2)  & = &
 e^2 c_0 
 \delta_{12} \left( \frac{\alpha_s}{2\pi} \right)^2 A^{(2)}_{2,\mbox{\tiny partial}}(q_1,\bar{q}_2).
\eq
There is only one independent helicity configuration, which we can
take to be
\bq
q_1^+,\bar{q}_2^-,p_3^-,p_4^+.
\eq
The helicity amplitude expressed in terms of spinorproducts
reads:
\bq
A^{(0)}_2(q_1^+,\bar{q}_2^-,p_3^-,p_4^+) =  
 2 i \frac{[ 14 ] \l 32 \r}{s_{12}}.
\eq
The leading colour contributions to $A^{(1)}_2$ and $A^{(2)}_2$ are:
\bq
A^{(1)}_{2,\mbox{\tiny partial}}(q_1^+,\bar{q}_2^-,p_3^-,p_4^+) & = & 
  \frac{1}{2} N_c V^{(1)}_2 A^{(0)}_2(q_1^+,\bar{q}_2^-,p_3^-,p_4^+)
  + O\left( \frac{1}{N_c^2} \right),
 \nonumber \\
A^{(2)}_{2,\mbox{\tiny partial}}(q_1^+,\bar{q}_2^-,p_3^-,p_4^+) & = & 
  \frac{1}{4} \left( N_c^2 -1 \right) V^{(2)}_2 A^{(0)}_2(q_1^+,\bar{q}_2^-,p_3^-,p_4^+)
  + O\left( \frac{1}{N_c} \right).
\eq
The leading-colour primitive amplitudes are therefore
\bq
A_2^{(1)} & = & V^{(1)}_2 A^{(0)}_2, \nonumber \\
A_2^{(2)} & = & V^{(2)}_2 A^{(0)}_2.
\eq
The renormalized coefficients $V^{(1)}_2$ and $V^{(2)}_2$ are given in
the HV scheme by 
%\cite{Matsuura:1988wt,Matsuura:1989sm,Kramer:1987sg,Kramer:1989sgerr,Giele:2002hx}
\cite{Matsuura:1988wt}-\cite{Giele:2002hx}
\bq
\lefteqn{
V^{(1)}_2  = 
} \nonumber \\
 & &
            -\frac{1}{\eps^2}
            -\frac{3}{2\eps}
            - 4 + \frac{7}{12} \pi^2
            + \left( -8 + \frac{7}{8} \pi^2 + \frac{7}{3} \zeta_3 \right) \eps
            + \left( -16 + \frac{7}{3} \pi^2 + \frac{7}{2} \zeta_3 
                     -\frac{73}{1440} \pi^4 \right) \eps^2
 \nonumber \\
 & &
            + i \pi \left[ 
                          -\frac{1}{\eps}
                          -\frac{3}{2}
                          +\left(-4 + \frac{\pi^2}{4} \right) \eps
                          + \left( -8 +\frac{3}{8}\pi^2 + \frac{7}{3} \zeta_3
                            \right) \eps^2
                    \right]
             + O(\eps^3),
 \nonumber \\
\lefteqn{
V^{(2)}_2 = 
} \nonumber \\
& &
           \frac{1}{2\eps^4}
         + \frac{17}{4 \eps^3}         
         + \left( \frac{433}{72} - \pi^2 \right) \frac{1}{\eps^2}
         + \left( \frac{4045}{432} - \frac{83}{24} \pi^2 + \frac{7}{6} \zeta_3 \right) \frac{1}{\eps}
         - \frac{9083}{2592} - \frac{2153}{432} \pi^2 
 \nonumber \\
 & & 
                  + \frac{26}{9} \zeta_3 
                  + \frac{263}{720} \pi^4 
             + O(\eps) + \mbox{imaginary parts}.
\eq
Note that for a NNLO calculation of $e^+ e^- \rightarrow \;\mbox{2 jets}$
one needs the real part of $V^{(1)}_2$ to order $O(\eps^2)$
and the imaginary part of $V^{(1)}_2$ to order $O(\eps)$.

% section ------------------------------------------------------------------

\subsection{Ultraviolet renormalization}

The amplitudes in the examples above are the renormalized ones, i.e. 
the ultraviolet subtraction has been performed.
To obtain the renormalized amplitudes in the $\overline{\mbox{MS}}$ scheme
from the bare ones, 
one replaces the bare coupling $\alpha_0$ with the renormalized coupling $\alpha_s(\mu^2)$ evaluated
at the renormalization scale $\mu^2$:
\begin{eqnarray}
\alpha_0 & = & \alpha_s S_\eps^{-1} \mu^{2\eps} \left[ 1 
 -\frac{\beta_0}{\eps} \left( \frac{\alpha_s}{2\pi} \right)
 + \left( \frac{\beta_0^2}{\eps^2} - \frac{\beta_1}{2\eps} \right) 
   \left( \frac{\alpha_s}{2\pi} \right)^2
 + {\cal O}(\alpha_s^3) \right],
\end{eqnarray}
where 
\begin{eqnarray}
S_\eps & = & \left( 4 \pi \right)^\eps e^{-\eps\gamma_E} \, ,
\end{eqnarray}
is the typical phase-space volume factor in $D =4-2\eps$ dimensions, 
$\gamma_E$ is Euler's constant,
and $\beta_0$ and $\beta_1$ are the first two coeffcients of the QCD $\beta$-function:
\begin{eqnarray}
\beta_0 = \frac{11}{6} C_A - \frac{2}{3} T_R N_f,
&\;\;\;&
\beta_1 = \frac{17}{6} C_A^2 - \frac{5}{3} C_A T_R N_f - C_F T_R N_f,
\end{eqnarray}
with the color factors
\begin{eqnarray}
C_A = N, \;\;\; C_F = \frac{N^2-1}{2N}, \;\;\; T_R = \frac{1}{2}.
\end{eqnarray}
Let the expansion in the strong coupling 
of the unrenormalized amplitude be
\bq
{\cal A}_{n,bare} & = & 
 \left( 4 \pi \alpha \right) \left( 4 \pi \alpha_0 \right)^{\frac{n-2}{2}}
 c_0 \; {\cal C} \; 
 \left[
        A_{n,bare}^{(0)} + \left( \frac{\alpha_0}{2\pi} \right) A_{n,bare}^{(1)}
                    + \left( \frac{\alpha_0}{2\pi} \right)^2 A_{n,bare}^{(2)}
        + O(\alpha_s^3)
 \right] 
 \nonumber \\
 & &
 + \; \mbox{subleading colour structures}.
\eq
where ${\cal C}$ is a colour factor.
Then, the renormalized two-loop amplitude can be expessed as
\bq
\lefteqn{
{\cal A}_{n,ren} = }
 \nonumber \\
 & & 
 \left( 4 \pi \alpha \right) \left( 4 \pi \alpha_s \right)^{\frac{n-2}{2}}
 c_0 \; {\cal C} \; \left( S_\eps^{-1} \mu^{2\eps} \right)^{\frac{n-2}{2}}
 \left[
        A_{n,ren}^{(0)} + \left( \frac{\alpha_s}{2\pi} \right) A_{n,ren}^{(1)}
                    + \left( \frac{\alpha_s}{2\pi} \right)^2 A_{n,ren}^{(2)}
        + O(\alpha_s^3)
 \right] 
  \nonumber \\
 & &
 + \; \mbox{subleading colour structures}.
\eq
The relations between
the renormalized and the bare amplitudes are
given by
\bq
A_{n,ren}^{(0)} & = & A_{n,bare}^{(0)},
 \nonumber \\
A_{n,ren}^{(1)} & = & S_\eps^{-1} \mu^{2\eps} A_{n,bare}^{(1)}
                      - \frac{(n-2)}{2} \frac{\beta_0}{\eps} A_{n,bare}^{(0)},
 \nonumber \\
A_{n,ren}^{(2)} & = & S_\eps^{-2} \mu^{4\eps} A_{n,bare}^{(2)}
                      - \frac{n}{2} \frac{\beta_0}{\eps} S_\eps^{-1} \mu^{2\eps} A_{n,bare}^{(1)}
                      + \frac{(n-2)}{2} 
                        \left( n \frac{\beta_0^2}{4\eps^2} - \frac{\beta_1}{2\eps} \right) A_{n,bare}^{(0)}.
\eq
Below I will discuss the factorization of one-loop amplitudes in singular
(soft and collinear) limits.
The general structure is 
\bq
\label{factsing}
A^{(1)}_{n}
  & = &
  \mbox{Sing}^{(0)} 
  \cdot A^{(1)}_{n-1} +
  \mbox{Sing}^{(1)} \cdot A^{(0)}_{n-1},
\eq
where the function $\mbox{Sing}$ corresponds either to the eikonal
factor (soft limit) or a splitting function (collinear limit).
Eq. (\ref{factsing}) holds both for bare and renormalized quantitites.
The relation between the renormalized and the bare expression
for the singular function is
\bq
\mbox{Sing}_{ren}^{(1)} & = & S_\eps^{-1} \mu^{2\eps} \; \mbox{Sing}_{bare}^{(1)}
                      - \frac{\beta_0}{2 \eps} \; \mbox{Sing}_{bare}^{(0)}.
\eq

% section ------------------------------------------------------------------

\section{Singular limits}
\label{sectlim}

In this section I review the behaviour of amplitudes in singular limits, e.g.
when two or three partons become degenerate.
The knowledge of the singular limits serves as a starting point
for the construction of the subtraction terms.
Most of the material in this section is summarized from the literature
\cite{Gehrmann-DeRidder:1998gf}-\cite{Kosower:2003cz},
although some results, like the triple collinear splitting function
for colour-ordered amplitudes, are new.
When considering singular limits it is convenient to work in a physical gauge.
In this case, only diagrams where the splitting occurs at external lines contribute
\cite{Ellis:1979ty,Catani:1999ss}.
In this paper I use the axial gauge $n \cdot A = 0$. In this 
gauge the gluon propagator reads
\bq
\frac{i}{k^2} d^{\mu\nu}(k,n) & = &
\frac{i}{k^2} \left( -g^{\mu\nu} + \frac{k^\mu n^\nu + n^\mu k^\nu}{k \cdot n}
 - n^2 \frac{k^\mu k^\nu}{\left( k \cdot n \right)^2}
\right)
\nonumber \\
& = & 
 - \frac{i}{k^2} \left( g^{\mu\rho} - \frac{k^\mu n^\rho}{k \cdot n} \right)
                 \left( g_\rho^{\;\;\;\nu} - \frac{n_\rho k^\nu }{k \cdot n} \right).
\eq

\subsection{Soft gluons}

If a single gluon becomes soft, the partial tree amplitude factorizes according to
\bq
A_n^{(0)}(p_1,p_2,p_3,...) & = & \mbox{Eik}^{(0,1)}(p_1,p_2,p_3) A_{n-1}^{(0)}(p_1,p_3,...),
\eq
where the eikonal factor is given by
% signs checked
\bq
\mbox{Eik}^{(0,1)}(p_1,p_2,p_3)
 & = & 
 4 \frac{p_1^\rho F_{\rho \mu \sigma}(p_2) p_3^\sigma}{s_{12} s_{23}}
 \eps^\mu(p_2)
\eq
and $F_{\rho \mu \sigma}(p)$ is defined by
\bq
F^{\rho \mu \sigma}(p) & = & g^{\rho \mu} p^\sigma - p^\rho g^{\mu \sigma}.
\eq
The square of the eikonal factor is given by
\bq
\left| \mbox{Eik}^{(0,1)}(p_1,p_2,p_3) \right|^2  & = & 
 4 \frac{s_{13}}{s_{12}s_{23}}.
\eq
In a similar way the partial tree amplitude factorizes, 
when two gluons become soft \cite{Berends:1989zn}
\bq
A_n^{(0)}(p_1,p_2,p_3,p_4,...) & = & \mbox{Eik}^{(0,2)}(p_1,p_2,p_3,p_4) A_{n-2}^{(0)}(p_1,p_4,...),
\eq
with
\bq
\mbox{Eik}^{(0,2)}(p_1,p_2,p_3,p_4)
 & = & 
 8  \left[ 
          \frac{p_1^\rho F_{\rho \mu \sigma}(p_2) F_{\sigma \nu \tau}(p_3) p_4^\tau}{s_{12} s_{23} s_{34}}
- \frac{p_1^\rho F_{\rho \mu \sigma}(p_2) F_{\sigma \nu \tau}(p_3) p_1^\tau}{s_{12} s_{23} (s_{12} + s_{13})}
 \right.
\nonumber \\
& &
 \left.
-  \frac{p_4^\rho F_{\rho \mu \sigma}(p_2) F_{\sigma \nu \tau}(p_3) p_4^\tau}{s_{23} s_{34} (s_{24} + s_{34})}
 \right]
  \eps^\mu(p_2) \eps^\nu(p_3).
\eq
The square of the double soft eikonal factor is given by
\bq
\lefteqn{
\left| \mbox{Eik}^{(0,2)}(p_1,p_2,p_3,p_4) \right|^2 
 = 
 8 \left[
          (1-\rho \eps) \frac{(s_{123} s_{34} + s_{12} s_{234} - s_{123} s_{234})^2}{s_{123}^2 s_{23}^2 s_{234}^2}
   \right.
}  \nonumber \\ 
   & & \left.
         + \frac{s_{14}^2}{s_{12} s_{34} s_{123} s_{234}}
         +\frac{s_{14}}{s_{23}}
          \left( \frac{1}{s_{12} s_{34}}
                +\frac{1}{s_{12} s_{234}}
                +\frac{1}{s_{123} s_{34}}
                -\frac{4}{s_{123} s_{234}}
           \right)
    \right].
\eq
I introduced the parameter $\rho$, which specifies the variant of dimensional
regularization:
$\rho  = 1$ for the CDR/HV schemes and $\rho=0$ for the FD scheme.
In the soft-gluon limit,
a one-loop primitive amplitude factorizes according to
\bq
A^{(1)}_{n}(p_1,p_2,p_3,...)
  & = &
  \mbox{Eik}^{(0,1)}(p_1,p_2,p_3) 
  \cdot A^{(1)}_{n-1}(p_1,p_3,...) \nonumber \\
& &
+
  \mbox{Eik}^{(1,1)}(p_1,p_2,p_3) \cdot A^{(0)}_{n-1}(p_1,p_3,...),
\eq
For the example considered here, the relevant 
one-loop eikonal function is given by
\cite{Bern:1999ry}
\bq
\lefteqn{
\mbox{Eik}^{(1,1)}(p_1,p_2,p_3)
 = }
 \nonumber \\
 & & 
 \left[
 - \frac{S_\eps^{-1} c_\Gamma}{\eps^2} 
 \Gamma(1+\eps) \Gamma(1-\eps)
 \left( \frac{\mu^2 (-s_{13})}{(-s_{12}) (-s_{23}) } \right)^\eps
 - \frac{11}{6\eps} 
 \right]
 \mbox{Eik}^{(0,1)}(p_1,p_2,p_3).
\eq
Here,
\bq 
c_\Gamma & = & (4\pi)^{\eps} \frac{\Gamma(1+\eps)\Gamma^2(1-\eps)}{\Gamma(1-2\eps)}.
\eq
As in the tree-level case with one soft gluon, the eikonal factor
is independent of the variant of the dimensional regularization scheme used.

\subsection{Collinear particles}

In the collinear limit tree amplitudes factorize according to
\bq
\label{factcollinearlimit}
A_n^{(0)}(...,p_i,p_j,...) & = & 
 \sum\limits_{\lambda} \; \mbox{Split}^{(0,1)}(p_i,p_j) \; A_{n-1}^{(0)}(...,p,...).
\eq
where $p_i$ and $p_j$ are the momenta of two adjacent legs and
the sum is over all polarizations.
In the collinear limit we parametrize the momenta of the partons $i$ and $j$ as
\cite{Catani:1997vz}
\bq
\label{collinearlimit}
p_i & = & z p + k_\perp - \frac{k_\perp^2}{z} \frac{n}{2 p n }, \nonumber \\
p_j & = & (1-z)  p - k_\perp - \frac{k_\perp^2}{1-z} \frac{n}{2 p n }.
\eq
Here $n$ is a massless four-vector and the transverse component $k_\perp$ satisfies
$2pk_\perp = 2n k_\perp =0$.
The collinear limits occurs for $k_\perp^2 \rightarrow 0$.
The splitting amplitudes $\mbox{Split}^{(0,1)}$ are universal, they depend
only on the two momenta becoming collinear, and not upon the specific amplitude under
consideration.
The splitting functions $\mbox{Split}^{(0,1)}$ are given by
\bq
% signs checked
% the splitting functions here are (-1) the ones in hep-ph/9903516
% and hep-ph/9903515
\mbox{Split}^{(0,1)}_{q \rightarrow q g} & = &
 \frac{1}{s_{ij}} \bar{u}(p_i) \eps\!\!\!/(p_j) u(p),
\nonumber \\
\mbox{Split}^{(0,1)}_{g \rightarrow g g} & = &
 \frac{2}{s_{ij}} \left[
    \eps(p_i) \cdot \eps(p_j) \; p_i \cdot \eps(p)
  + \eps(p_j) \cdot \eps(p) \; p_j \cdot \eps(p_i)
  - \eps(p_i) \cdot \eps(p) \; p_i \cdot \eps(p_j)
 \right],
\nonumber \\
\mbox{Split}^{(0,1)}_{g \rightarrow q \bar{q}} & = &
 \frac{1}{s_{ij}} \bar{u}(p_i) \eps\!\!\!/(p) u(p_j).
\eq
Due to the sum over spins in eq. (\ref{factcollinearlimit}),
spin correlations are retained in the collinear limit of squared tree amplitudes.
With 
\bq
P^{(0,1)} & = &  \sum\limits_{\lambda, \lambda'}
 u(p) \left. \; \mbox{Split}^{(0,1)} \right.^\ast \mbox{Split}^{(0,1)} \;\bar{u}(p)
 \;\;\;\;\mbox{for quarks,}
 \nonumber \\
P^{(0,1)} & = &  \sum\limits_{\lambda, \lambda'}
 \left. \eps^\mu(p) \right.^\ast \;
                \left. \mbox{Split}^{(0,1)} \right.^\ast \mbox{Split}^{(0,1)} \; \eps^\nu(p)
 \;\;\;\;\mbox{for gluons,}
\eq
and the paramterization eq. (\ref{collinearlimit}) one finds
\bq
 P^{(0,1)}_{q \rightarrow q g } & = & 
   \frac{2}{s_{ij}} p\!\!\!/ \left[ \frac{2z}{1-z} + (1 - \rho \eps) (1-z) \right], \nonumber \\
 P^{(0,1)}_{g \rightarrow g g} & = & 
   \frac{2}{s_{ij}}  \left[ - g^{\mu\nu} \left( \frac{2z}{1-z} + \frac{2(1-z)}{z} \right) 
   - 4 (1-\rho \eps) z (1-z) \frac{k^\mu_\perp k^\nu_\perp}{k_\perp^2} \right], \nonumber \\
 P^{(0,1)}_{g \rightarrow q \bar{q}} & = & 
   \frac{2}{s_{ij}} \left[ -g^{\mu\nu} + 4 z (1-z) \frac{k^\mu_\perp k^\nu_\perp}{k_\perp^2} \right].
\eq
One-loop primitive amplitudes factorize according to
%\cite{Bern:1994zx,Bern:1998sc,Kosower:1999xi,Kosower:1999rx,Bern:1999ry}
\cite{Bern:1994zx}-\cite{Bern:1999ry}
\bq
\label{factloopcollinearlimit}
\lefteqn{
A_n^{(1)}(...,p_i,p_j,...) = }
 \nonumber \\
 & & 
 \sum\limits_{\lambda} \; \mbox{Split}^{(0,1)}(p_i,p_j) \; A_{n-1}^{(1)}(...,p,...)
 +
 \sum\limits_{\lambda} \; \mbox{Split}^{(1,1)}(p_i,p_j) \; A_{n-1}^{(0)}(...,p,...).
\eq
Here, the splitting amplitudes $\mbox{Split}^{(1,1)}$ occur as a new structure.
For the one-loop amplitude $A_{3}^{(1)}$ for $e^+ e^- \rightarrow q g \bar{q}$
we only need the $q \rightarrow q g$ splitting function.
This function is given by
\cite{Kosower:1999rx,Bern:1999ry}
% signs checked
% notes the minus sign in front of f2
\bq
\mbox{Split}^{(1,1)}_{q \rightarrow q g} & = &
  S_\eps^{-1} c_\Gamma \left( \frac{-s_{12}}{\mu^2} \right)^{-\eps}
     \left( (f_1(z)+f_2) \; \mbox{Split}^{(0,1)}_{q \rightarrow q g}
           - f_2 \; \frac{2p_i \cdot \eps(p_j)}{s_{ij}^2} 
                 \bar{u}(p_i) p\!\!\!/_j u(p)
     \right)
 \nonumber \\
 & &
  - \frac{11}{6\eps} \; \mbox{Split}^{(0,1)}_{q \rightarrow q g},
\eq
where the coefficients $f_1(z)$ and $f_2$ are
\bq
f_1(z) & = &
 -\frac{1}{\eps^2} \; {}_2F_1\left(1,-\eps,1-\eps;\frac{-z}{1-z}\right),
 \nonumber \\
f_2 & = & \frac{1-\rho\eps}{2(1-\eps)(1-2\eps)}.
\eq
The $\eps$-expansion of the hypergeometric function is
\bq
{}_2F_1\left(1,-\eps,1-\eps;\frac{-z}{1-z}\right)
 & = & 1 - \sum\limits_{k=1}^{\infty} \eps^k 
           \; \mbox{Li}_k\left(\frac{-z}{1-z}\right).
\eq 
The hypergeometric function can also be written as
\bq
{}_2F_1\left(1,-\eps,1-\eps;\frac{-z}{1-z}\right)
 & = &
 \Gamma(1+\eps) \Gamma(1-\eps) \left( \frac{z}{1-z} \right)^\eps 
 + 1 - z^\eps \; {}_2F_1\left(\eps,\eps,1+\eps;1-z\right),
 \nonumber \\
\eq
which is useful when studying the soft limit.
The interference term of $\mbox{Split}^{(0,1)}$ with $\mbox{Split}^{(1,1)}$
is given by
\bq
\label{oneloopcollinterference}
\lefteqn{
P^{(1,1)}_{q \rightarrow q g} =
 \sum\limits_{\lambda}
 u(p) \left( \mbox{Split}^{(0,1)}_{q \rightarrow q g} \right)^\ast 
 \mbox{Split}^{(1,1)}_{q \rightarrow q g} \;\bar{u}(p) \; + c.c.
 = }
 \nonumber \\
 & &
  S_\eps^{-1} c_\Gamma \left( \frac{-s_{12}}{\mu^2} \right)^{-\eps}
     \left\{ f_1(z) \; P^{(0,1)}_{q \rightarrow q g}
           + f_2 \; \frac{2}{s_{ij}} p\!\!\!/   
                 \left[ 1 - \rho \eps (1-z) \right]
     \right\} 
  - \frac{11}{6\eps} P^{(0,1)}_{q \rightarrow q g} \; + c.c.
\eq
Here ``c.c.'' denotes the complex conjugate.
\\
\\
For tree amplitudes
we also have to consider the case, where three partons become collinear
simultaneously.
In the triple collinear limit we parametrize the momenta as follows
\cite{Catani:1999ss}:
\bq
\label{triplecollmom}
p_i & = & z_i p + k_{\perp i} - \frac{k_{\perp i}^2}{z_i} \frac{n}{2 p n},
\eq
where $n^2 = 0$, $2 k_{\perp i} p = 2 k_{\perp i} n = 0$,
and the $z_i$ and $k_{\perp i}$ satisfy
\bq
\sum z_i = 1, & & \sum k_{\perp i} = 0.
\eq
The following formulae are useful:
\bq
z_i =  \frac{2 p_i n }{2 p n} = \frac{2 p_i n }{2 p_{ijk} n},
 \;\;\;
s_{ij} = - z_i z_j \left( \frac{k_{\perp i}}{z_i} - \frac{k_{\perp j}}{z_j} \right)^2,
 \;\;\;
k_{\perp i}^2  = z_i \left( s_{jk} - (1-z_i) s_{ijk} \right).
 & &
\eq
Here, $p_{ijk} = p_i + p_j + p_k$.
For the example of the leading $N_c$ contributions to
$e^+ e^- \rightarrow q g g \bar{q}$ we have to consider only the triple
collinear limit $q \rightarrow q g g$.
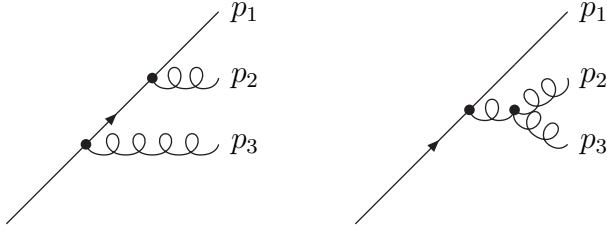
\begin{figure}
\begin{center}
\begin{tabular}{cc}
\begin{picture}(120,100)(0,0)
 \ArrowLine(10,10)(90,90)
 \Vertex(40,40){2}
 \Vertex(65,65){2}
 \Gluon(40,40)(90,40){-4}{4}
 \Gluon(65,65)(90,65){-4}{2}
 \Text(95,90)[l]{$p_1$}
 \Text(95,65)[l]{$p_2$}
 \Text(95,40)[l]{$p_3$}
\end{picture}
&
\begin{picture}(120,100)(0,0)
 \ArrowLine(10,10)(70,70)
 \Line(70,70)(90,90)
 \Vertex(53,53){2}
 \Gluon(53,53)(70,53){-4}{1}
 \Vertex(70,53){2}
 \Gluon(70,53)(90,65){-4}{2}
 \Gluon(70,53)(90,40){-4}{2}
 \Text(95,90)[l]{$p_1$}
 \Text(95,65)[l]{$p_2$}
 \Text(95,40)[l]{$p_3$}
\end{picture}
\\
\end{tabular}
\caption{\label{triplecollinear} Diagrams contributing to the triple collinear limit
$q \rightarrow q g g$.}
\end{center}
\end{figure}
In the triple collinear limit the tree amplitude factorizes according to
\bq
\label{facttriplecollinearlimit}
A_n^{(0)}(p_1,p_2,p_3,...) & = & 
 \sum\limits_{\lambda} \; \mbox{Split}^{(0,2)}_{q\rightarrow qgg}(p_1,p_2,p_3) \; A_{n-2}^{(0)}(p,...).
\eq
Fig. (\ref{triplecollinear}) shows the Feynman diagrams contributing to the triple
collinear limit $q \rightarrow q g g$.
We are interested in the square of the splitting function
$\mbox{Split}^{(0,2)}_{q\rightarrow qgg}$, defined by
\bq
P^{(0,2)} & = &  \sum\limits_{\lambda, \lambda'}
 u(p) \left. \; \mbox{Split}^{(0,2)} \right.^\ast \mbox{Split}^{(0,2)} \;\bar{u}(p)
 \;\;\;\;\mbox{for quarks.}
\eq
For the splitting $q \rightarrow q g g$ I find
\bq
\label{Pqtoqgg}
\lefteqn{
P^{(0,2)}_{q\rightarrow qgg} = }
\nonumber \\
 & &
 \frac{4}{s_{123}^2} p\!\!\!/
 \left\{
       2(1-\rho\eps) \frac{\left( z_3 - (1-x_1) z_{23} \right)^2}{x_2^2 z_{23}^2}
       + 4 (1-\rho\eps) \frac{x_1}{x_2}
       + (1-\rho\eps)^2 \frac{x_2}{x_1}
       + 3 - 4\rho\eps + \rho^2\eps^2
 \right. \nonumber \\
 & & \left. 
       + \frac{1}{x_1 x_2} 
         \left[
               \frac{1+(1-z_3)^2-\rho\eps z_3^2}{z_{23}}
              +\frac{1+(1-z_{23})^2-\rho\eps z_{23}^2}{z_3}
         \right]
 \right. \nonumber \\
 & & \left. 
       + \frac{1}{x_1}
         \left[
               \frac{2}{z_3 z_{23}} - \frac{2}{z_3} - \frac{2}{z_{23}} 
               - \frac{1+(1-z_3)^2-\rho\eps z_3^2}{z_{23}}
               + (1-\rho\eps) \frac{z_3}{z_{23}}
               + (1-\rho\eps) \frac{z_{23}}{z_{3}}
 \right. \right. \nonumber \\
 & & \left. \left. 
               + \rho\eps (1-\rho\eps) (z_{23}-z_3)
               - 1 + 3 \rho\eps
         \right]
 \right. \nonumber \\
 & & \left. 
       + \frac{1}{x_2}
         \left[
               -2 \frac{1+(1-z_3)^2-\rho\eps z_3^2}{z_{23}}
               +\frac{1+(1-z_{23})^2-\rho\eps z_{23}^2}{z_3}
               -\frac{4}{z_{23}}
               - 4\rho\eps \frac{z_3}{z_{23}}
 \right. \right. \nonumber \\
 & & \left. \left. 
               - 3 (1-\rho\eps) (z_{23}-z_3)
               + 2 + 4 \rho\eps
         \right]
 \right\}.
\eq
Here $x_1=s_{12}/s_{123}$, $x_2=s_{23}/s_{123}$ and $z_{23} = z_2 + z_3$.
Note that eq. (\ref{Pqtoqgg}) gives the triple collinear splitting
function for colour-ordered amplitudes. This implies a fixed cyclic
ordering for the two gluons. 
In contrast, the triple collinear splitting functions
given in ref. \cite{Campbell:1998hg} and ref. \cite{Catani:1999ss} 
are the ones for the full matrix
element squared, e.g. they include a sum over the permutations
of the two gluons.

\subsection{Factorizable double unresolved contributions}

For the double unresolved contributions there are also two configurations,
which have quite a simple factorizable structure in the singular limit.
These two cases are: a) the emission of two independent pairs of collinear particles
and b) the emission of a soft gluon together with one pair of collinear particles.
Let us first consider the case of the emission of two independent pairs of
collinear particles.
I consider the case $p_1||p_2$ and $p_3||p_4$. The amplitude factorizes according
to
\bq
\label{paircoll}
A_n^{(0)}(p_1,p_2,p_3,p_4) & = & 
 \sum\limits_{\lambda,\lambda'} \; \mbox{Split}^{(0,1)}(p_1,p_2) 
                                \; \mbox{Split}^{(0,1)}(p_3,p_4) \; A_{n-2}^{(0)}(p,q),
\eq
where $p=p_1+p_2+O(k_{\perp}^2)$ and $q=p_3+p_4+O(k_{\perp}^2)$. 
For the leading-colour contributions to $e+ e- \rightarrow q g g \bar{q}$
only the splittings $q \rightarrow q g$ and $\bar{q} \rightarrow \bar{q} g$ are
relevant.
Squaring the splitting functions in eq. (\ref{paircoll}) one obtains
\bq
P^{(0,2)}_{q\rightarrow qg,\bar{q} \rightarrow \bar{q} g} & = &
 P^{(0,1)}_{q\rightarrow qg}
 \; P^{(0,1)}_{\bar{q} \rightarrow \bar{q} g}.
\eq
In the case of the emission of a soft gluon together with a pair of 
collinear partons it is for our purposes sufficient to consider
the case of a soft gluon ($p_3$) together with a collinear
splitting $q \rightarrow q g$ of the partons $p_1$ and $p_2$.
I simply quote the singular factor for the squared matrix element 
\cite{Campbell:1998hg,Catani:1999ss}:
\bq
P^{(0,2)}_{q\rightarrow q_1g_2,g_3 \;\mbox{\tiny soft}} & = &
 4 \frac{s_{p4}}{s_{p3}s_{34}} \; P^{(0,1)}_{q\rightarrow qg}.
\eq
Here, $p=p_1+p_2+O(k_{\perp}^2)$ and the singular factor factorizes
into an eikonal factor times a splitting function.

% section ------------------------------------------------------------------

\section{Subtraction terms}
\label{sectsubtr}

In this section I first review the subtraction terms for single unresolved contributions.
I then give the subtraction terms for double unresolved contributions
and for contributions with one virtual and one real unresolved parton.

\subsection{Subtraction terms for single unresolved contributions}

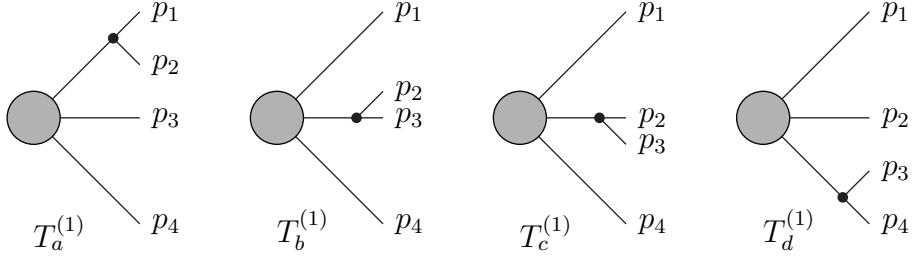
\begin{figure}
\begin{center}
\begin{tabular}{cccc}
\begin{picture}(80,100)(0,0)
 \Line(20,50)(60,90)
 \Line(20,50)(60,50)
 \Line(20,50)(60,10)
 \Vertex(50,80){2}
 \Line(50,80)(60,70)
 \GCirc(20,50){10}{0.7}
 \Text(30,0)[b]{$T^{(1)}_a$}
 \Text(65,90)[l]{$p_1$}
 \Text(65,70)[l]{$p_2$}
 \Text(65,50)[l]{$p_3$}
 \Text(65,10)[l]{$p_4$}
\end{picture}
&
\begin{picture}(80,100)(0,0)
 \Line(20,50)(60,90)
 \Line(20,50)(60,50)
 \Line(20,50)(60,10)
 \Vertex(50,50){2}
 \Line(50,50)(60,60)
 \GCirc(20,50){10}{0.7}
 \Text(30,0)[b]{$T^{(1)}_b$}
 \Text(65,90)[l]{$p_1$}
 \Text(65,60)[l]{$p_2$}
 \Text(65,50)[l]{$p_3$}
 \Text(65,10)[l]{$p_4$}
\end{picture}
&
\begin{picture}(80,100)(0,0)
 \Line(20,50)(60,90)
 \Line(20,50)(60,50)
 \Line(20,50)(60,10)
 \Vertex(50,50){2}
 \Line(50,50)(60,40)
 \GCirc(20,50){10}{0.7}
 \Text(30,0)[b]{$T^{(1)}_c$}
 \Text(65,90)[l]{$p_1$}
 \Text(65,50)[l]{$p_2$}
 \Text(65,40)[l]{$p_3$}
 \Text(65,10)[l]{$p_4$}
\end{picture}
&
\begin{picture}(80,100)(0,0)
 \Line(20,50)(60,90)
 \Line(20,50)(60,50)
 \Line(20,50)(60,10)
 \Vertex(50,20){2}
 \Line(50,20)(60,30)
 \GCirc(20,50){10}{0.7}
 \Text(30,0)[b]{$T^{(1)}_d$}
 \Text(65,90)[l]{$p_1$}
 \Text(65,50)[l]{$p_2$}
 \Text(65,30)[l]{$p_3$}
 \Text(65,10)[l]{$p_4$}
\end{picture}
\\
\end{tabular}
\caption{\label{nlofig} Splitting topologies for NLO subtraction.
At each splitting the emitted particle is directed towards the spectator.}
\end{center}
\end{figure}

The approxiamtion term $d \alpha^{(0,1)}_{n+1}$ is given as a sum over dipoles \cite{Catani:1997vz}:
\bq
d \alpha^{(0,1)}_{n+1} & = & 
 \sum\limits_{\mbox{\tiny topologies $T$}} {\cal D}_{n+1}^{(0,1)}(T).
\eq
For leading colour contributions the sum reduces to combinations
where the emitting parton $i$ and the emitted parton $j$ are adjacent, 
and the spectator $k$ is either adjacent to $i$ or $j$.
It is also convenient to denote by $k'$ the other parton adjacent to the pair
$i$ and $j$.
Therefore the cyclic ordering is either $k',i,j,k$ or $k,j,i,k'$.
The relevant topologies for the example of the leading $N_c$ contributions to
$e^+ e^- \rightarrow q g g \bar{q}$ are shown in fig. (\ref{nlofig}).
A dipol is constructed from amputated amplitudes, evaluated with mapped momenta
and dipole splitting functions.
By an amputated amplitude I mean an amplitude where a polarization vector 
(or external spinor for fermions) has been removed.
I denote amputated amplitudes by $\left| A \right\r$.
If the amputated parton is a gluon, one has
\bq
\left| A_n^{(0)}(...,p,...) \right\r
 & = & 
 \frac{\partial}{\partial \eps_\mu(p)} A_n^{(0)}(...,p,...). 
\eq
If the amputated parton is a quark, one has
\bq
\left| A_n^{(0)}(...,p,...) \right\r
 & = & 
 \frac{\partial}{\partial \bar{u}(p)} A_n^{(0)}(...,p,...).
\eq
The dipole subtraction terms are obtained by sandwiching a dipole splitting function
in between these amputated amplitudes:
\bq
{\cal D}_{n+1}^{(0,1)}(T) & = &
 c \; \left\l A_{n+1}^{(0)}(...,p_e,...) \right|
 {\cal P}^{(0,1)}(T)
 \left| A_{n+1}^{(0)}(...,p_e,...) \right\r d\phi_{n+2}.
\eq
Here, $c$ denotes an overall prefactor. 
For the example $e^+ e^- \rightarrow q g g \bar{q}$, it is given by
\bq
c & = &
 \left( 4 \pi \alpha \right)^2 \left( 4 \pi \alpha_s \right)^2 
 \left| c_0 \right|^2
 \times 
 \frac{1}{4} \left( N_c^2 -1 \right) N_c.
\eq
The dipole splitting functions read
\bq
\label{nlosplittingfcts}
{\cal P}^{(0,1)}_{q \rightarrow q g}(p_k',p_i,p_j,p_k) & = & 
   \frac{2}{s_{ijk}} \frac{1}{y} p\!\!\!/_{e} 
         \left[ \frac{2}{1-z(1-y)} - 2  + \left( 1 - \rho \eps \right) (1-y) (1-z) \right],
\nonumber \\
{\cal P}^{(0,1)}_{g \rightarrow g g}(p_k',p_i,p_j,p_k) & = & 
   \frac{2}{s_{ijk}} \frac{1}{y}  
         \left[ - g^{\mu\nu} \left( \frac{2}{1-z(1-y)} - 2 \right)
           + \left( 1 - \rho \eps \right) \frac{4 r (1-y)^2}{s_{ij}} k_\perp^\mu k_\perp^\nu \right],
\nonumber \\
{\cal P}^{(0,1)}_{g \rightarrow q \bar{q}}(p_k',p_i,p_j,p_k) & = & 
   \frac{2}{s_{ijk}} \frac{1}{y}  
         \left[ - \frac{1}{2} g^{\mu\nu} 
           - \frac{4 r (1-y)^2}{s_{ij}} k_\perp^\mu k_\perp^\nu \right],
\eq
where
\bq
r & = & \frac{2p_{k'}p_e}{2p_{k'}p_e+2p_e p_s}.
\eq
These splitting functions differ by finite terms from the original choice of
Catani and Seymour. 
The new set of splitting functions in eq.(\ref{nlosplittingfcts}) is introduced
for later convenience with the NNLO subtraction terms.
Although arbitrary non-singular terms may be added to the NLO subtraction terms,
the NNLO subtraction terms depend on the actual form of the NLO subtraction terms.

The dipole splitting functions in eq. (\ref{nlosplittingfcts}) 
have also the correct singular behaviour 
in the soft limit. The term $1/(1-z(1-y))$ equals $s_{ijk}/(s_{ij}+s_{jk})$
and combining this term with the one from the dipole, where
the r\^oles of emitter and spectator are exchanged, one obtains
\bq
\frac{1}{s_{ij}} \frac{s_{ijk}}{\left(s_{ij}+s_{jk}\right)}
+\frac{1}{s_{jk}} \frac{s_{ijk}}{\left(s_{ij}+s_{jk}\right)}
 & = & 
 \frac{s_{ijk}}{s_{ij} s_{jk}},
\eq
which is the eikonal factor in the soft limit.

For the $q \rightarrow q g$ splitting, the quark polarization
matrix $p\!\!\!/_{e}$ is included in the definition of the dipole splitting function.
For the $g \rightarrow g g$ and $g \rightarrow q \bar{q}$ splittings,
the full splitting function is shared between the dipoles with
spectators $k$ and $k'$, where $k$ and $k'$ are the partons adjacent
to the $(i,j)$ pair in the colour ordered 
amplitude $A^{(0)}(...,k',i,j,k,...)$.
For the $g \rightarrow g g$ splitting the dipole splitting function
is chosen in such a way that the singularity when parton $j$ becomes
soft resides in ${\cal P}_{g \rightarrow g g}(p_i,p_j;p_k)$, where as the singularity when parton $i$
becomes soft resides in ${\cal P}_{g \rightarrow g g}(p_j,p_i;p_{k'})$.
The amputated amplitudes are evaluated with mapped momenta.
The mapping of momenta 
relates a $n+1$ parton configuration to
a $n$ parton configuration:
\bq
P_{n+1 \rightarrow n} & : & \left(p_1,...,p_{n+1}\right) \rightarrow \left(p_1',...,p_{n}'\right).
\eq 
Such a mapping has to satisfy momentum conservation and the on-mass-shell conditions.
Furthermore it must have the right behaviour in the singular limits.
Several choices for such a mapping exist \cite{Catani:1997vz,Kosower:1998zr}.
One possible choice relates three parton momenta of the $n+1$ parton configuration
to two parton momenta of the $n$ parton configuration, while leaving
the remaining $n-2$ parton momenta unchanged.
This mapping is given by
\bq
\label{reconstructCataniSeymour}
p_e & = & p_i + p_j - \frac{y}{1-y} p_k, 
\nonumber \\
p_s & = & \frac{1}{1-y} p_k.
\eq
$p_i$, $p_j$ and $p_k$ are the momenta of the $n+1$ parton configuration,
and $p_e$ and $p_s$ are the resulting momenta of the of the $n$ parton configuration.
The variables $y$ and $z$ are given by
\bq
y = \frac{s_{ij}}{s_{ijk}}, & & z = \frac{s_{ik}}{s_{ik}+s_{jk}}.
\eq
It is easily verified that this mapping satisfies momentum conservation
\bq
p_e + p_s & = & p_i + p_j + p_k,
\eq
and the on-shell conditions
\bq
p_e^2 = 0, & & p_s^2 = 0.
\eq
The singular limit corresponds to $y \rightarrow 0$ and in this limit we have
\bq
p_e - (p_i+p_j) = O(y), 
& & 
p_s - p_k = O(y).
\eq
It is also useful to introduce the vector $k_\perp$, defined by
\bq
k_\perp & = & (1-z) p_i - z p_j - \left( 1 -2 z \right) \frac{y}{1-y} p_k.
\eq
With this vector and the value of the longitudinal momentum fraction $z$, 
the mapping may be inverted to yield
\bq
p_i & = & z p_e + k_\perp + y \left( 1 - z \right) p_s, \nonumber \\
p_j & = & (1-z) p_e - k_\perp + y z p_s, \nonumber \\
p_k & = & (1-y) p_s,
\eq
where $y$ is now given by
\bq
y & = & \frac{-k_\perp^2}{z(1-z) \left(p_e+p_s\right)^2}.
\eq
The singular limit occurs now for $k_\perp \rightarrow 0$.
The $g \rightarrow g g$ and $g \rightarrow q \bar{q}$ splittings also
involve spin correlation through the spin correlation tensor $k_\perp^\mu k_\perp^\nu$.
Note that the spin correlation tensor $k_\perp^\mu k_\perp^\nu$
is orthogonal to $p_e$ and $p_s$:
\bq
2 p_e k_\perp = 2 p_s k_\perp = 0.
\eq
Further note that the contraction of $p_e$ into an amputated amplitude vanishes
due to gauge invariance.
\\
The parameter $\rho$ specifies the variant of dimensional regularization:
$\rho = 1$ corresponds to the CDR/HV schemes and $\rho=0$ to a four-dimensional
scheme.

\subsection{Subtraction terms for double unresolved contributions}

In this subsection I give the subtraction terms for double
unresolved contributions.
To be more precise, these subtraction terms have to approximate
\bq
\label{singleanddoubleunres}
{\cal O}_{n+2} \; d\sigma_{n+2}^{(0)} 
             - {\cal O}_{n+1} \circ d\alpha^{(0,1)}_{n+1}
\eq
in all single and double unresolved limits.
Eq. (\ref{singleanddoubleunres}) is integrable over single unresolved
regions for observables, which vanish on $n$-parton configurations,
e.g. $(n+1)$-jet observables.
However, as discussed in the example of eq. (\ref{divsingleunres}),
for a generic $n$-jet observable, eq. (\ref{singleanddoubleunres})
is in general not integrable over single unresolved regions.
Therefore, not only the singularities corresponding to double unresolved
limits, but also the singularities corresponding to single unresolved
limits have to be subtracted.
\\
\\
The subtraction term $d\alpha^{(0,2)}_{n}$ at NNLO is also written
as a sum over topologies:
\bq
d \alpha^{(0,2)}_{n} & = & 
 \sum\limits_{\mbox{\tiny topologies $T$}} {\cal D}_{n}^{(0,2)}(T).
\eq
These topologies should be thought of as pictorial representations
of the pole structure of the subtraction terms ${\cal D}_{n}^{(0,2)}(T)$
and the way the momenta of the $n$-parton configurations are constructed
from the momenta of the $(n+2)$-parton configurations.
The intuitive picture of ``iterated dipoles'' is 
only justified for the approximations of $d\alpha^{(0,1)}_{n+1}$.
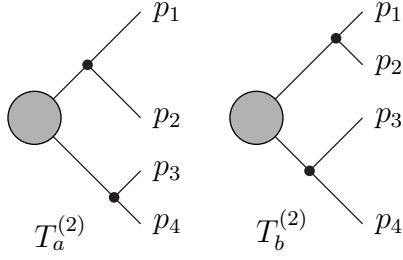
\begin{figure}
\begin{center}
\begin{tabular}{cc}
\begin{picture}(80,100)(0,0)
 \Line(20,50)(60,90)
 \Line(20,50)(60,10)
 \Vertex(40,70){2}
 \Line(40,70)(60,50)
 \Vertex(50,20){2}
 \Line(50,20)(60,30)
 \GCirc(20,50){10}{0.7}
 \Text(30,0)[b]{$T^{(2)}_\topoa$}
 \Text(65,90)[l]{$p_1$}
 \Text(65,50)[l]{$p_2$}
 \Text(65,30)[l]{$p_3$}
 \Text(65,10)[l]{$p_4$}
\end{picture}
\begin{picture}(80,100)(0,0)
 \Line(20,50)(60,90)
 \Line(20,50)(60,10)
 \Vertex(40,30){2}
 \Line(40,30)(60,50)
 \Vertex(50,80){2}
 \Line(50,80)(60,70)
 \GCirc(20,50){10}{0.7}
 \Text(30,0)[b]{$T^{(2)}_\topob$}
 \Text(65,90)[l]{$p_1$}
 \Text(65,70)[l]{$p_2$}
 \Text(65,50)[l]{$p_3$}
 \Text(65,10)[l]{$p_4$}
\end{picture}
\\
& \\
\end{tabular}
\caption{\label{nnlofig2} Splitting topologies for NNLO subtraction:
Two single splittings from two different branches.
At each splitting the emitted particle is directed towards the spectator.
}
\end{center}
\end{figure}
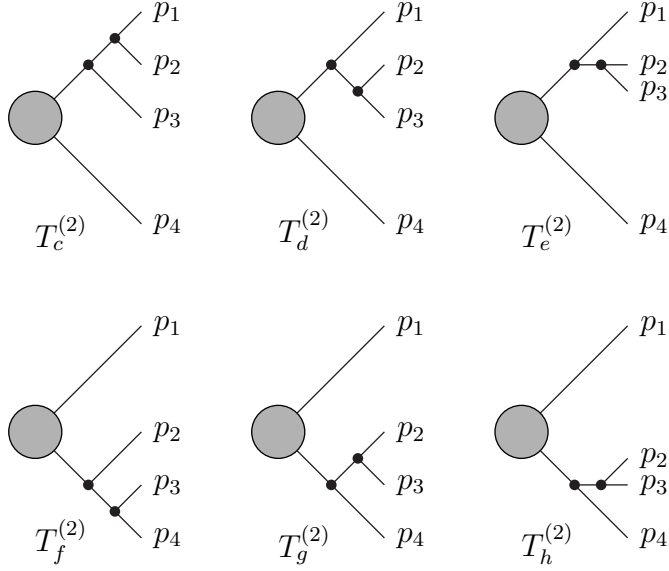
\begin{figure}
\begin{center}
\begin{tabular}{ccc}
\begin{picture}(80,100)(0,0)
 \Line(20,50)(60,90)
 \Line(20,50)(60,10)
 \Vertex(40,70){2}
 \Line(40,70)(60,50)
 \Vertex(50,80){2}
 \Line(50,80)(60,70)
 \GCirc(20,50){10}{0.7}
 \Text(30,0)[b]{$T^{(2)}_\topoc$}
 \Text(65,90)[l]{$p_1$}
 \Text(65,70)[l]{$p_2$}
 \Text(65,50)[l]{$p_3$}
 \Text(65,10)[l]{$p_4$}
\end{picture}
&
\begin{picture}(80,100)(0,0)
 \Line(20,50)(60,90)
 \Line(20,50)(60,10)
 \Vertex(40,70){2}
 \Line(40,70)(60,50)
 \Vertex(50,60){2}
 \Line(50,60)(60,70)
 \GCirc(20,50){10}{0.7}
 \Text(30,0)[b]{$T^{(2)}_\topod$}
 \Text(65,90)[l]{$p_1$}
 \Text(65,70)[l]{$p_2$}
 \Text(65,50)[l]{$p_3$}
 \Text(65,10)[l]{$p_4$}
\end{picture}
&
\begin{picture}(80,100)(0,0)
 \Line(20,50)(60,90)
 \Line(20,50)(60,10)
 \Vertex(40,70){2}
 \Line(40,70)(60,70)
 \Vertex(50,70){2}
 \Line(50,70)(60,60)
 \GCirc(20,50){10}{0.7}
 \Text(30,0)[b]{$T^{(2)}_\topoe$}
 \Text(65,90)[l]{$p_1$}
 \Text(65,70)[l]{$p_2$}
 \Text(65,60)[l]{$p_3$}
 \Text(65,10)[l]{$p_4$}
\end{picture}
\\
& & \\
\begin{picture}(80,100)(0,0)
 \Line(20,50)(60,90)
 \Line(20,50)(60,10)
 \Vertex(40,30){2}
 \Line(40,30)(60,50)
 \Vertex(50,20){2}
 \Line(50,20)(60,30)
 \GCirc(20,50){10}{0.7}
 \Text(30,0)[b]{$T^{(2)}_\topof$}
 \Text(65,90)[l]{$p_1$}
 \Text(65,50)[l]{$p_2$}
 \Text(65,30)[l]{$p_3$}
 \Text(65,10)[l]{$p_4$}
\end{picture}
&
\begin{picture}(80,100)(0,0)
 \Line(20,50)(60,90)
 \Line(20,50)(60,10)
 \Vertex(40,30){2}
 \Line(40,30)(60,50)
 \Vertex(50,40){2}
 \Line(50,40)(60,30)
 \GCirc(20,50){10}{0.7}
 \Text(30,0)[b]{$T^{(2)}_\topog$}
 \Text(65,90)[l]{$p_1$}
 \Text(65,50)[l]{$p_2$}
 \Text(65,30)[l]{$p_3$}
 \Text(65,10)[l]{$p_4$}
\end{picture}
&
\begin{picture}(80,100)(0,0)
 \Line(20,50)(60,90)
 \Line(20,50)(60,10)
 \Vertex(40,30){2}
 \Line(40,30)(60,30)
 \Vertex(50,30){2}
 \Line(50,30)(60,40)
 \GCirc(20,50){10}{0.7}
 \Text(30,0)[b]{$T^{(2)}_\topoh$}
 \Text(65,90)[l]{$p_1$}
 \Text(65,40)[l]{$p_2$}
 \Text(65,30)[l]{$p_3$}
 \Text(65,10)[l]{$p_4$}
\end{picture}
\\
& & \\
\end{tabular}
\caption{\label{nnlofig3} Splitting topologies for NNLO subtraction:
Two sequential splittings from the same branch.
At each splitting the emitted particle is directed towards the spectator.
}
\end{center}
\end{figure}

The relevant topologies for the leading $N_c$ contributions to
$e^+ e^- \rightarrow q g g \bar{q}$ are shown in fig. (\ref{nnlofig2}) to fig. (\ref{nnlofig3}).
They fall into two classes: 
two single splittings from two different branches
(topologies $T^{(2)}_\topoa$ and $T^{(2)}_\topob$)
and two sequential splittings from the same branch 
(topologies $T^{(2)}_\topoc$- $T^{(2)}_\topoh$).
It is sufficient to consider topologies $T^{(2)}_\topoa$ and
$T^{(2)}_\topoc$-$T^{(2)}_\topoe$ only. 
The remaining topologies are related by symmetry to the ones mentioned
above.
The subtraction term is given by
\bq
\label{gendoubleunres}
{\cal D}_{n}^{(0,2)}(T) & = &
 c \; {\cal J}(T) \; 
 \left\l A_{n}^{(0)}(...,p_e^{(2)},...,p_s^{(2)},...) \right|
 {\cal P}^{(0,2)}(T)
 \left| A_{n}^{(0)}(...,p_e^{(2)},...,p_s^{(2)},...) \right\r d\phi_{n+2}.
 \nonumber \\
\eq
${\cal J}$ is a prefactor not singular in any limit, which can be used
to absorbe Jacobians into the definition of the subtraction terms, thus rendering
the analytical integration over the unresolved phase space simpler.
Here, I simply take ${\cal J}=1$.
The splitting function ${\cal P}^{(0,2)}(T)$ is sandwiched between double-amputated amplitudes
for the topologies $T^{(2)}_\topoa$ and $T^{(2)}_\topob$, and is sandwiched between single-amputated
amplitudes for the topologies $T^{(2)}_\topoc$-$T^{(2)}_\topoh$.
It is convenient to write
\bq
 {\cal P}^{(0,2)}(T) & = & {\cal P}^{(0,2)}_{(0,0)}(T) - {\cal P}^{(0,2)}_{(0,1)}(T),
\eq
where ${\cal P}^{(0,2)}_{(0,0)}$ is an approxiamation to $d\sigma_{n+2}^{(0)}$
and ${\cal P}^{(0,2)}_{(0,1)}$ is an approxiamation to $d\alpha_{n+1}^{(0,1)}$.
The notation with super- and subscripts is as follows:
\bq
 {\cal P}^{(l,k)}_{(l',k')}(T)
\eq
contributes to $d\alpha^{(l,k)}$ and approximates a term of $d\alpha^{(l',k')}$,
(with $d\sigma^{(l)}$ identified as $d\alpha^{(l,0)}$).
It should be noted that the subtraction terms are not unique.
For example, the complete triple collinear splitting function 
$P_{q \rightarrow q g g}$ is shared between the topologies
$T^{(2)}_\topoc$, $T^{(2)}_\topod$ and $T^{(2)}_\topoe$.
Therefore, a term of the form
\bq
\frac{c}{s_{123}^2},
\eq
which is only singular in the triple collinear limit and finite in all other
cases, can be freely moved between these topologies.
On the other hand, a term of the form
\bq
\frac{c'}{s_{12} s_{123}},
\eq
which is singular in the triple collinear limit and in the single unresolved
limit $s_{12}\rightarrow 0$ is uniquely associated with
topology $T^{(2)}_\topoc$, as the pictorial representation suggests.
Below I will give one possible set for the subtraction terms.

I briefly mention how the $n$-parton momentum configuration is constructed from
the $(n+2)$-parton configuration.
As an example I discuss topology $T^{(2)}_\topoc$.
For the outermost splitting,
particle $1$ is the emitter, particle $2$ is the
emitted soft or collinear particle and particle $3$ is the spectator.
Particle $4$ is not involved in this splitting.
From this $4$-parton momentum configuration $(p_1,p_2,p_3,p_4)$ 
a $3$-parton momentum configuration $(p_e^{(1)},p_s^{(1)},p_4)$
is first constructed, using the Catani-Seymour reconstruction function
given in eq. (\ref{reconstructCataniSeymour}).
From the $3$-parton momentum configuration a $2$-parton momentum
configuration $(p_e^{(2)},p_s^{(2)})$ is then constructed, using the symmetric reconstruction
functions given by Kosower \cite{Kosower:1998zr}.
This symmetric choice is necessary to ensure that topologies
$T^{(2)}_\topoc$ and $T^{(2)}_\topob$ are evaluated with the same
$n$-parton momentum configuration.
This is necessary for a cancellation of (subleading) single
unresolved singularities in $s_{12} \rightarrow 0$ and is discussed
in detail in sect. \ref{sectcanc}.

To present the splitting functions for double unresolved configurations I use the following
notation:
\bq
\bar{y} = 1- y, \;\;\; \bar{z} = 1- z,
\eq
and $p_{ij}=p_i+p_j$.

\subsubsection{Topology \topoA}

This topology corresponds to two independent single splittings and the splitting
function ${\cal P}$ is inserted into a double-amputated amplitude.
The splitting functions read:
\bq
\lefteqn{
{\cal P}^{(0,2)}_{(0,0)\;q\rightarrow qg, \bar{q} \rightarrow \bar{q} g}\left(T^{(2)}_\topoa\right) = 
  \left( p\!\!\!/_e^{(2)} \right)_{ij}
  \left( p\!\!\!/_s^{(2)} \right)_{\bar{i}\bar{j}}
  \frac{4}{s_{1234}^2}
   \frac{1}{y_1 x_2 \left( \bar{y}_1 x_1 + y_1 x_2 \right) }
} \nonumber \\
 & & 
  \times 
  \left\{
    \left[ 1 + \frac{\bar{y}_1 \bar{z}_1}{u} \right] 
          \frac{2}{\bar{y}_1 x_1 + \left( y_1 + \bar{y}_1 \bar{z}_1 \right) x_2} 
   + \left[ \frac{1}{y_1 x_2 + u} + \frac{1}{y_1 + \bar{y}_1 \bar{z}_1} \right]
     \left[ -2 + (1-\rho \eps) x_2 \right]
 \right. \nonumber \\ & & \left.
   + \left[ \frac{1}{\bar{y}_1 x_1 + x_2} + \frac{u}{\bar{y}_1 x_1 + \bar{y}_1 \bar{z}_1 x_2} \right]
     \left[ -2 + (1-\rho \eps) u \right]
   + \left[ -2 + (1-\rho \eps) x_2 \right] \left[ -2 + (1-\rho \eps) u \right]
 \right. \nonumber \\ & & \left.
   - \rho \eps \left( \bar{y}_1 x_1 + y_1 x_2 \right)
          \left[ \frac{2}{y_1+\bar{y}_1 \bar{z}_1 } - 2 + (1-\rho \eps) \bar{y}_1 \bar{z}_1 \right] 
  \right\},
 \nonumber \\
\lefteqn{
{\cal P}^{(0,2)}_{(0,1)\;q\rightarrow qg, \bar{q} \rightarrow \bar{q} g}\left(T^{(2)}_\topoa\right) = 
  \left( p\!\!\!/_e^{(2)} \right)_{ij}
  \left( p\!\!\!/_s^{(2)} \right)_{\bar{i}\bar{j}}
  \frac{4}{s_{1234}^2}
   \frac{1}{x_1 x_2}
        \left[
               \frac{2}{x_1+x_2} - 2 + \left( 1- \rho \eps \right) x_2 
               - \rho \eps x_1
        \right]
} \nonumber \\
 & &
  \times \frac{1}{y_1}
        \left[
               \frac{2}{1-z_1 (1 - y_1)} - 2 + \left( 1 - \rho \eps \right) (1-y_1) (1-z_1) 
        \right].
\eq
As already noted, this is inserted into a double-amputated amplitude.
The variables $x_1$, $x_2$ and $x_3$ are given by
\bq
x_1 = y_2, \;\;\; x_2 = (1-y_2) (1-z_2), \;\;\; x_3 = (1-y_2) z_2. &&
\eq
The variables $y_1$, $z_1$, $y_2$ and $z_2$ are given by
\bq
y_1 = \frac{s_{34}}{s_{234}},
& &
z_1 = \frac{s_{24}}{s_{23}+s_{24}},
\nonumber \\
y_2 = \frac{1}{(1-y_1)} \frac{s_{12}}{s_{1234}},
& &
z_2 = 1 - \frac{1}{(1-y_2)} \frac{s_{234}}{s_{1234}}.
\eq
The remaining variable $u$ is given by
\bq
u & = & \left( z_1 + \bar{y}_1 \bar{z}_1 \right) x_1 + \bar{y}_1 \bar{z}_1 x_2 + \bar{z}_1 x_3
        + 2 (1-2w) \sqrt{y_1 z_1 \bar{z}_1 x_1 x_3}, \nonumber \\
w & = & \frac{1}{2} \left( 1 - c_2 \right), \nonumber \\
c_2 & = & \frac{ s_{12}s_{23}^2 
               + s_{24} 
                 \left( (s_{23}+s_{24})(s_{13}+s_{14})-s_{12}s_{23} \right) 
               - s_{14} (s_{23}+s_{24})^2
               }{ 2 s_{23} \sqrt{s_{12} s_{24}
                  \left( (s_{23}+s_{24})(s_{13}+s_{14})-s_{12}s_{23} \right)
                 }}.
\eq
The amputated amplitude is evaluated with the mapped momenta
\bq
\label{reconstructA}
p_e^{(2)} & = & 
 \frac{1}{2} \left[ 1 + q + \frac{\bar{y}_2 \bar{z}_2 \left( 1+ q - 2 r \right) }{1-\bar{y}_2 \bar{z}_2} 
                      \right] p_1
 + r p_s^{(1)}
 + \frac{1}{2} \left[ 1- q + \frac{y_2 \left( 1- q - 2 r \right) }{1-y_2} \right] p_e^{(1)},
 \nonumber \\
p_s^{(2)} & = &
 \frac{1}{2} \left[ 1 - q - \frac{\bar{y}_2 \bar{z}_2 \left( 1+ q - 2 r \right) }{1-\bar{y}_2 \bar{z}_2} 
                      \right] p_1
 + (1-r) p_s^{(1)}
 + \frac{1}{2} \left[ 1+ q - \frac{y_2 \left( 1- q - 2 r \right) }{1-y_2} \right] p_e^{(1)},
 \nonumber \\
\eq
where
\bq
p_e^{(1)} = p_{34} - \frac{y_1}{1-y_1} p_2, 
 & &
p_s^{(1)} = \frac{1}{1-y_1} p_2,
\eq
and the variables $r$ and $q$ are given by
\bq
r = \frac{\bar{y}_2 \bar{z}_2}{y_2+\bar{y}_2 \bar{z}_2},
& &
q = \sqrt{1+ 4 r (1-r) y_2 \frac{1-z_2}{z_2}}.
\eq

\subsubsection{Topology \topoC}

The splitting functions read:
\bq
\lefteqn{
{\cal P}^{(0,2)}_{(0,0)\; q\rightarrow q g g}\left(T^{(2)}_\topoc\right) 
  =  
        p\!\!\!/_e^{(2)}
        \frac{4}{s_{1234}^2} 
        \frac{1}{x_1^2} 
 \left\{
  \frac{1}{y_1 \left( y_1 + \bar{y}_1 \bar{z}_1 \right)}
    \left[
           \frac{2}{\left(y_1 + \bar{y}_1 \bar{z}_1 \right) x_1 + \bar{y}_1 x_2}
          + \frac{2}{y_1 x_1 +u }
 \right. \right. } \nonumber \\ & &\left. \left.
          + \frac{u}{\left(y_1 + \bar{y}_1 \bar{z}_1 \right) x_1 + \bar{y}_1 x_2}
            \left( - 2 + (1-\rho \eps) u  \right) 
          + \frac{\bar{y}_1 x_2}{u} 
            \left( -2 + (1-\rho \eps ) \bar{y}_1 x_2 \right) 
    \right]
 \right. \nonumber \\ & &\left. 
  + \frac{1}{y_1}
    \left[
            \frac{2}{u \left[ \left(y_1 + \bar{y}_1 \bar{z}_1 \right)x_1 
                              + \bar{y}_1 x_2 \right]}
           + \frac{1}{x_1 + \bar{y}_1 x_2}
             \left( -2 + (1-\rho \eps) u \right)
           - \frac{4}{u}
           + (1-\rho \eps) \frac{\bar{y}_1 x_2}{u}
 \right. \right. \nonumber \\ & &\left. \left.
           - \frac{\bar{y}_1 x_2}{u} \left( -2 + (1-\rho \eps) \bar{y}_1 x_2 \right)
           + \rho \eps ( 1- \rho \eps) u
           - \rho \eps ( 1- \rho \eps) \bar{y}_1 x_2
           - 1 + 3 \rho \eps 
    \right]
 \right. \nonumber \\ & &\left. 
    + \left( 1 - \rho \eps \right)^2 \frac{\bar{y}_1 \bar{z}_1}{y_1}
    + 3 - 4 \rho \eps + \rho^2 \eps^2
    + \frac{x_1}{y_1}
      \left[
             - \rho \eps \left( \frac{2}{y_1+\bar{y}_1 \bar{z}_1} 
                                 -2 + (1-\rho \eps) \bar{y}_1 \bar{z}_1 \right)
             + \frac{2}{u}
 \right. \right. \nonumber \\ & &\left. \left.
             - (1-\rho \eps) \left( \frac{x_1}{u} - 2 \bar{y}_1 z_1 
                                    + (1-\rho\eps)  \bar{y}_1 z_1 x_1 \right)
    + \frac{1}{u \left( y_1 + \bar{y}_1 \bar{z}_1 \right)}
          \left( -2 + (1-\rho \eps) x_1 \right)
      \right] 
 \right\},
 \nonumber \\
\lefteqn{
{\cal P}^{(0,2)}_{(0,1)\; q\rightarrow q g g}\left(T^{(2)}_\topoc\right) 
  =  
        p\!\!\!/_e^{(2)}
        \frac{4}{s_{1234}^2} 
        \frac{1}{x_1^2} 
        \left[
               \frac{2}{x_1+x_2} - 2 + \left( 1- \rho \eps \right) x_2 
               - \rho \eps x_1
        \right]
}
  \nonumber \\
  & & \times
        \frac{1}{y_1}  
        \left[
               \frac{2}{1-z_1 (1 - y_1)} - 2 + \left( 1 - \rho \eps \right) (1-y_1) (1-z_1) 
        \right].
\eq
The variables $x_1$, $x_2$ and $x_3$ are given by
\bq
x_1 = y_2, \;\;\; x_2 = (1-y_2) (1-z_2), \;\;\; x_3 = (1-y_2) z_2. &&
\eq
The variables $y_1$, $z_1$, $y_2$ and $z_2$ are given by
\bq
y_1 = \frac{s_{12}}{s_{123}},
& &
z_1 = \frac{s_{13}}{s_{13}+s_{23}},
\nonumber \\
y_2 = \frac{s_{123}}{s_{1234}},
& &
z_2 = 1 - \frac{1}{(1-y_1)(1-y_2)} \frac{s_{34}}{s_{1234}}.
\eq
The remaining variable $u$ is given by
\bq
u & = & \bar{y}_1 \bar{z}_1 x_1 + \left( z_1 + \bar{y}_1 \bar{z}_1 \right) x_2 + \bar{z}_1 x_3
        + 2 (1-2w) \sqrt{y_1 z_1 \bar{z}_1 x_2 x_3}, \nonumber \\
w & = & \frac{1}{2} \left( 1 - c_2 \right), \nonumber \\
c_2 & = & \frac{ s_{12}s_{23}s_{34} 
               + s_{13} 
                 \left( (s_{13}+s_{23})(s_{14}+s_{24})-s_{12}s_{34} \right) 
               - s_{14} (s_{13}+s_{23})^2
               }{ 2 \sqrt{s_{12} s_{23} s_{34} s_{13}
                 \left( (s_{13}+s_{23})(s_{14}+s_{24})-s_{12}s_{34} \right) 
                 }}.
\eq
The amputated amplitude is evaluated with the mapped momenta
\bq
\label{reconstructC}
p_e^{(2)} & = &
 \frac{1}{2} \left[ 1+ q - \frac{\bar{y}_2 \bar{z}_2 \left( 1- q - 2 r \right) }{1-\bar{y}_2 \bar{z}_2} \right] p_e^{(1)}
 + (1-r) p_s^{(1)}
 +\frac{1}{2} \left[ 1 - q - \frac{y_2\left( 1+ q - 2 r \right) }{1-y_2 } 
                      \right] p_4,
 \nonumber \\
p_s^{(2)} & = & 
 \frac{1}{2} \left[ 1- q + \frac{\bar{y}_2 \bar{z}_2 \left( 1- q - 2 r \right) }{1-\bar{y}_2 \bar{z}_2} \right] p_e^{(1)}
 + r p_s^{(1)}
 + \frac{1}{2} \left[ 1 + q + \frac{y_2 \left( 1+ q - 2 r \right) }{1-y_2} 
                      \right] p_4,
 \nonumber \\
\eq
where
\bq
p_e^{(1)} = p_{12} - \frac{y_1}{1-y_1} p_3, 
 & &
p_s^{(1)} = \frac{1}{1-y_1} p_3,
\eq
and the variables $r$ and $q$ are given by
\bq
r = \frac{y_2}{y_2+\bar{y}_2 \bar{z}_2},
& &
q = \sqrt{1+ 4 r (1-r) y_2 \frac{1-z_2}{z_2}}.
\eq

\subsubsection{Topology \topoD}

The splitting functions read:
\bq
\lefteqn{
{\cal P}^{(0,2)}_{(0,0)\; q\rightarrow q g g}\left(T^{(2)}_\topod\right) 
  =  
    p\!\!\!/_e^{(2)}
    \frac{4}{s_{1234}^2}
    \frac{1}{x_1^2}  
    \left\{
      \frac{1}{y_1\left( y_1+\bar{y}_1\bar{z}_1\right)}
      \left[
       \frac{2}{\left( y_1+\bar{y}_1\bar{z}_1\right)x_1+u}
 \right. \right.
 } \nonumber \\
 & & \left. \left.
      + \frac{2}{\left( y_1+\bar{y}_1\bar{z}_1\right) x_1+x_2+y_1x_3}
      + \frac{y_1 x_1 + x_2 + y_1 x_3}{\left( y_1+\bar{y}_1\bar{z}_1\right) x_1 +u }
        \left[ -2 + (1-\rho \eps) \left(y_1 x_1 + x_2 + y_1 x_3\right) \right]
 \right. \right. \nonumber \\ && \left. \left.
      + \frac{u}{y_1 x_1 + x_2 + y_1 x_3}
        \left[ -2 + (1-\rho \eps) u \right]
     \right]
 \right. \nonumber \\ && \left.
   + \frac{1}{y_1}
     \left[
           - \frac{4}{(1+y_1)x_1 + x_2 +y_1 x_3}
           + \frac{2}{x_1+u}
           - \frac{2u}{y_1 x_1 + x_2 + y_1 x_3} \left[ -2 + (1-\rho \eps) u\right]
 \right. \right. \nonumber \\ && \left. \left.
           - 4 \rho \eps \frac{u}{y_1 x_1 + x_2 + y_1 x_3}
           - (1-\rho \eps) \left(y_1 x_1 + x_2 + y_1 x_3\right)
           + 3 (1-\rho \eps) u
           - 2
           + 4 \rho \eps
     \right]
 \right. \nonumber \\ && \left.
   + 4 (1-\rho \eps) \frac{\bar{y}_1\bar{z}_1}{y_1}
   + \frac{x_1}{y_1} ( - \rho \eps)
     \left[ \frac{2}{y_1+\bar{y}_1\bar{z}_1} -2 + \bar{y}_1^2 z_1 \bar{z}_1 \right]
 \right. \nonumber \\ && \left.
   + \frac{2(1-\rho \eps)}{y_1^2 \left[(1+y_1) x_1 + x_2 + y_1 x_3\right]}
     \left[
              \left( 1- \bar{y}_1 \bar{z}_1 \right)^2 
              \left(y_1 x_1 + x_2 + y_1 x_3\right)
             - 2 u \left( 1- \bar{y}_1 \bar{z}_1 \right)
 \right. \right. \nonumber \\ && \left. \left.
             + \frac{u^2}{y_1 x_1 + x_2 + y_1 x_3}
      \right]
   \right\},
 \nonumber \\
\lefteqn{
{\cal P}^{(0,2)}_{(0,1)\; q\rightarrow q g g}\left(T^{(2)}_\topod\right) 
  =  
        p\!\!\!/_e^{(2)}
        \frac{4}{s_{1234}^2}
    \left\{
        \frac{1}{x_1^2}  
         \left[
               \frac{2}{x_1+x_2} - 2 + \left( 1- \rho \eps \right) x_2 
               - \rho \eps x_1
         \right]
 \right. } \nonumber \\
 & & \left. 
  \times
        \frac{1}{y_1}  
        \left[
               \frac{2}{1-z_1 (1 - y_1)} - 2  
        \right]
       + \frac{8 \left( 1 - \rho \eps \right) x_3}{x_1^2 \left(x_1+x_2\right)}
                ( 1- 2w)^2  
         \frac{1}{y_1} (1-y_1)^2 z_1 (1-z_1)
 \right. \nonumber \\ & & \left. 
       + \frac{2}{x_1+x_2}
         \left[
                \left( 1- \rho \eps  \right) \left( \frac{x_2^2}{x_1^2} + 1 \right) 
                - 2 \rho \eps \frac{x_2}{x_1}
         \right]
         \frac{1}{y_1} (1-y_1)^2 z_1 (1-z_1)
    \right\}.
\eq
The variables $x_1$, $x_2$ and $x_3$ are given by
\bq
x_1 = y_2, \;\;\; x_2 = (1-y_2) (1-z_2), \;\;\; x_3 = (1-y_2) z_2. &&
\eq
The variables $y_1$, $z_1$, $y_2$ and $z_2$ are given by
\bq
y_1 = \frac{s_{23}}{s_{123}},
& &
z_1 = \frac{s_{13}}{s_{12}+s_{13}},
\nonumber \\
y_2 = \frac{s_{123}}{s_{1234}},
& &
z_2 = \frac{1}{(1-y_1)(1-y_2)} \frac{s_{14}}{s_{1234}}.
\eq
The remaining variable $u$ is given by
\bq
u & = & z_1 x_2 + y_1 \bar{z}_1 x_3
        - 2 (1-2w) \sqrt{y_1 z_1 \bar{z}_1 x_2 x_3}, \nonumber \\
w & = & \frac{1}{2} \left( 1 - c_2 \right), \nonumber \\
c_2 & = & \frac{ s_{12}s_{23}s_{14} 
               + s_{13} 
                 \left( (s_{12}+s_{13})(s_{24}+s_{34})-s_{23}s_{14} \right) 
               - s_{34} (s_{12}+s_{13})^2
               }{ 2 \sqrt{s_{12} s_{23} s_{13} s_{14}
                 \left( (s_{12}+s_{13})(s_{24}+s_{34})-s_{23}s_{14} \right) 
                 }}.
\eq
The amputated amplitude is evaluated with the mapped momenta
\bq
\label{reconstructD}
p_e^{(2)} & = &
 \frac{1}{2} \left[ 1+ q - \frac{\bar{y}_2 \bar{z}_2 \left( 1- q - 2 r \right) }{1-\bar{y}_2 \bar{z}_2} \right] p_s^{(1)}
 + (1-r) p_e^{(1)}
 + \frac{1}{2} \left[ 1 - q - \frac{y_2 \left( 1+ q - 2 r \right) }{1-y_2} 
                      \right] p_4,
 \nonumber \\
p_s^{(2)} & = & 
 \frac{1}{2} \left[ 1- q + \frac{\bar{y}_2 \bar{z}_2 \left( 1- q - 2 r \right) }{1-\bar{y}_2 \bar{z}_2} \right] p_s^{(1)}
 + r p_e^{(1)}
 + \frac{1}{2} \left[ 1 + q + \frac{y_2 \left( 1+ q - 2 r \right) }{1-y_2} 
                      \right] p_4,
 \nonumber \\
\eq
where
\bq
p_e^{(1)} = p_{23} - \frac{y_1}{1-y_1} p_1, 
 & &
p_s^{(1)} = \frac{1}{1-y_1} p_1,
\eq
and the variables $r$ and $q$ are given by
\bq
r = \frac{y_2 }{y_2+\bar{y}_2 \bar{z}_2},
& &
q = \sqrt{1+ 4 r (1-r) y_2 \frac{1-z_2}{z_2}}.
\eq

\subsubsection{Topology \topoE}

The splitting functions read:
\bq
\lefteqn{
{\cal P}^{(0,2)}_{(0,0)\; q\rightarrow q g g}\left(T^{(2)}_\topoe\right) 
  =  
        p\!\!\!/_e^{(2)}
        \frac{4}{s_{1234}^2} 
        \left\{
        \frac{1}{(x_1+y_1x_2+y_1x_3)y_1 x_2} 
  \right. } \nonumber \\
 & & \left.
 \times
        \left[
             - \frac{4}{x_1+\left(1+y_1\right)x_2+y_1x_3}
             + \frac{1}{\left( y_1 + \bar{y}_1\bar{z}_1 \right)} \left[-2+(1-\rho\eps)x_2 \right]
             + 4 - 2 (1-\rho\eps) x_2
        \right]
 \right. \nonumber \\ 
 & & \left.
 + \frac{1}{y_1 x_2 \left[ \left(y_1 + \bar{y}_1\bar{z}_1 \right) x_2 + u \right]}
   \frac{1}{(y_1 + \bar{y}_1\bar{z}_1)}
       \left[ -2 + (1-\rho \eps) x_2 \right]
 \right. \nonumber \\ 
 & & \left.
 + \frac{1}{y_1 x_2} (-\rho \eps)
        \left[
                \frac{2}{y_1+\bar{y}_1 \bar{z}_1} -2 + \bar{y}_1^2 z_1 \bar{z}_1
        \right]
 \right\},
\nonumber \\
\lefteqn{
{\cal P}^{(0,2)}_{(0,1)\; q\rightarrow q g g}\left(T^{(2)}_\topoe\right) 
  =  
        p\!\!\!/_e^{(2)}
        \frac{4}{s_{1234}^2} 
        \frac{1}{x_1 x_2}
        \left[
               \frac{2}{x_1+x_2} - 2 +\left( 1 - \rho \eps \right) x_2 
               - \rho \eps x_1
        \right]
 } \nonumber \\
 & &
 \times
        \frac{1}{y_1}  
        \left[
               \frac{2}{1-z_1 (1 - y_1)} - 2 
        \right].
 \nonumber \\
\eq
The variables $x_1$, $x_2$ and $x_3$ are given by
\bq
x_1 = y_2, \;\;\; x_2 = (1-y_2) (1-z_2), \;\;\; x_3 = (1-y_2) z_2. &&
\eq
The variables $y_1$, $z_1$, $y_2$ and $z_2$ are given by
\bq
y_1 = \frac{s_{23}}{s_{234}},
& &
z_1 = \frac{s_{24}}{s_{24}+s_{34}},
\nonumber \\
y_2 = \frac{1}{(1-y_1)} 
      \frac{(s_{123} s_{234} - s_{23} s_{1234})}{s_{234} s_{1234}},
& &
z_2 = 1 - \frac{1}{(1-y_2)} \frac{s_{234}}{s_{1234}}.
\eq
The remaining variable $u$ is given by
\bq
u & = & z_1 x_1 + y_1 \bar{z}_1 x_3
        - 2 (1-2w) \sqrt{y_1 z_1 \bar{z}_1 x_1 x_3}, \nonumber \\
w & = & \frac{1}{2} \left( 1 - c_2 \right), \nonumber \\
c_2 & = & \frac{ s_{23}s_{34}s_{14} 
               + s_{24} 
                 \left( (s_{12}+s_{13})(s_{24}+s_{34})-s_{23}s_{14} \right) 
               - s_{12} (s_{24}+s_{34})^2
               }{ 2 \sqrt{s_{23} s_{34} s_{14} s_{24}
                 \left( (s_{12}+s_{13})(s_{24}+s_{34})-s_{23}s_{14} \right) 
                 }}.
\eq
The amputated amplitude is evaluated with the mapped momenta
\bq
\label{reconstructE}
p_e^{(2)} & = & 
 \frac{1}{2} \left[ 1 + q + \frac{\bar{y}_2 \bar{z}_2 \left( 1+ q - 2 r \right) }{1-\bar{y}_2 \bar{z}_2} 
                      \right] p_1
 + r p_e^{(1)}
 + \frac{1}{2} \left[ 1- q + \frac{y_2 \left( 1- q - 2 r \right) }{1-y_2} \right] p_s^{(1)},
 \nonumber \\
p_s^{(2)} & = &
 \frac{1}{2} \left[ 1 - q - \frac{\bar{y}_2 \bar{z}_2 \left( 1+ q - 2 r \right) }{1-\bar{y}_2 \bar{z}_2} 
                      \right] p_1
 + (1-r) p_e^{(1)}
 + \frac{1}{2} \left[ 1+ q - \frac{y_2 \left( 1- q - 2 r \right) }{1-y_2} \right] p_s^{(1)},
 \nonumber \\
\eq
where
\bq
p_e^{(1)} = p_{23} - \frac{y_1}{1-y_1} p_4, 
 & &
p_s^{(1)} = \frac{1}{1-y_1} p_4,
\eq
and the variables $r$ and $q$ are given by
\bq
r = \frac{\bar{y}_2 \bar{z}_2}{y_2+\bar{y}_2 \bar{z}_2},
& &
q = \sqrt{1+ 4 r (1-r) y_2 \frac{1-z_2}{z_2}}.
\eq

\subsection{Subtraction terms for contributions with one virtual and one real unresolved parton}

\begin{figure}
\begin{center}
\begin{tabular}{cc}
\begin{picture}(80,100)(0,0)
 \Line(20,50)(60,90)
 \Line(20,50)(60,10)
 \Vertex(40,70){2}
 \Line(40,70)(60,50)
 \GCirc(20,50){10}{0.7}
 \Text(30,0)[b]{$T^{(1,1)}_a$}
 \Text(65,90)[l]{$p_1$}
 \Text(65,50)[l]{$p_2$}
 \Text(65,10)[l]{$p_3$}
\end{picture}
&
\begin{picture}(80,100)(0,0)
 \Line(20,50)(60,90)
 \Line(20,50)(60,10)
 \Vertex(40,30){2}
 \Line(40,30)(60,50)
 \GCirc(20,50){10}{0.7}
 \Text(30,0)[b]{$T^{(1,1)}_b$}
 \Text(65,90)[l]{$p_1$}
 \Text(65,50)[l]{$p_2$}
 \Text(65,10)[l]{$p_3$}
\end{picture}
\\
& \\
\end{tabular}
\caption{\label{nnlofig4} Splitting topologies for the subtraction terms
for the one-loop amplitude with one unresolved parton.
}
\end{center}
\end{figure}
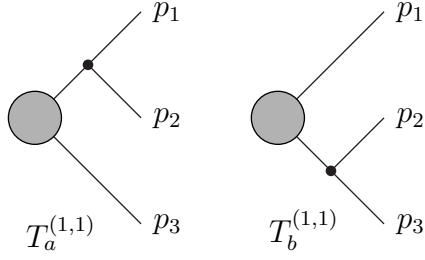

In this section I give the subtracttion term $d\alpha_n^{(1,1)}$,
which approximates
\bq
d\sigma_{n+1}^{(1)} + d\alpha_{n+1}^{(0,1)}
 & = & 
  \left( 4 \pi \alpha \right)^2 \left( 4 \pi \alpha_s \right)
  \frac{\alpha_s}{2\pi} 
 \left| c_0 \right|^2
 \times 
 \frac{1}{4} \left( N_c^2 -1 \right) N_c 
 \nonumber \\
 & &
 \times
 \left[
         \left. A_{n+1}^{(0)} \right.^\ast A_{n+1}^{(1)}
       + \left. A_{n+1}^{(1)} \right.^\ast A_{n+1}^{(0)}
       + \left. A_{n+1}^{(0)} \right.^\ast {\cal I} A_{n+1}^{(0)}
 \right]
 d\phi_{n+1}
\eq 
${\cal I}$ is obtained from integrating $d\alpha_{n+1}^{(0,1)}$ 
over the unresolved phase space and is given for our example by
\bq
{\cal I} & = & S_\eps^{-1} \frac{(4 \pi)^\eps}{\Gamma(1-\eps)}
 \left\{ 
        \left( \frac{s_{12}}{\mu^2} \right)^{-\eps} 
          \left[ {\cal V}^{(0,1)}_{q \rightarrow q g} 
                +{\cal V}^{(0,1)}_{g \rightarrow g g}\left(\frac{s_{23}}{s_{12}+s_{23}}\right)
          \right]
 \right. \nonumber \\
 & & \left. 
      + \left( \frac{s_{23}}{\mu^2} \right)^{-\eps} 
          \left[ {\cal V}^{(0,1)}_{q \rightarrow q g} 
                +{\cal V}^{(0,1)}_{g \rightarrow g g}\left(\frac{s_{12}}{s_{12}+s_{23}}\right)
          \right]
 \right\}.
\eq
The explicit forms of the 
functions ${\cal V}^{(0,1)}_{q \rightarrow q g}$ and 
${\cal V}^{(0,1)}_{q \rightarrow q g}$ are given in eq. (\ref{ressingunresolved2}).
The subtraction term $d\alpha^{(1,1)}_{n}$ is written
as a sum over topologies:
\bq
d \alpha^{(1,1)}_{n} & = & 
 \sum\limits_{\mbox{\tiny topologies $T$}} {\cal D}_{n}^{(1,1)}(T).
\eq
The relevant topologies for the leading $N_c$ contributions to
$e^+ e^- \rightarrow q g g \bar{q}$ are shown in fig. (\ref{nnlofig4}).
It is sufficient to consider the topology $T^{(1,1)}_a$.
The topology $T^{(1,1)}_b$ is related by symmetry to $T^{(1,1)}_a$.
The subtraction term is given by
\bq
\lefteqn{
{\cal D}_{n}^{(1,1)}(T) = 
} \\
 & &
 \frac{c}{8\pi^2}
  \left[
         \left\l A_{n}^{(0)} \right| {\cal P}^{(0,1)}(T) \left| A_{n}^{(1)} \right\r 
       + \left\l A_{n}^{(1)} \right| {\cal P}^{(0,1)}(T) \left| A_{n}^{(0)} \right\r 
       + \left\l A_{n}^{(0)} \right| {\cal P}^{(1,1)}(T) \left| A_{n}^{(0)} \right\r 
  \right] d\phi_{n+1}.
 \nonumber 
\eq
Here, in the first and second term 
the NLO splitting function ${\cal P}^{(0,1)}$ is sandwiched between
$A_{n}^{(0)}$ and $A_{n}^{(1)}$.
The third term contains as a new structure the one-loop splitting function
${\cal P}^{(1,1)}$.
It is convenient to write
\bq
 {\cal P}^{(1,1)}(T) & = & {\cal P}^{(1,1)}_{(1,0)}(T) + {\cal P}^{(1,1)}_{(0,1)}(T),
\eq
where ${\cal P}^{(1,1)}_{(1,0)}$ is (part of ) the approxiamation to $d\sigma_{n+1}^{(1)}$
and ${\cal P}^{(1,1)}_{(0,1)}$ is an approxiamation to $d\alpha_{n+1}^{(0,1)}$.
The splitting functions read:
\bq
\label{oneloopsplitsubtrterm}
\lefteqn{
{\cal P}^{(1,1)}_{(1,0)\; q \rightarrow q g}\left(T^{(1,1)}_a \right) 
 =  
 }
 \nonumber \\
 & & 
 S_\eps^{-1} c_\Gamma \left( \frac{-s_{123}}{\mu^2} \right)^{-\eps} 
  y^{-\eps}
     \left\{ 
         g_1(y,z) \; {\cal P}^{(0,1)}_{q \rightarrow q g}
         + g_2 \frac{2}{s_{123}} \frac{1}{y} p\!\!\!/_{e} 
               \left[ 1 - \rho \eps (1-y) (1-z) \right]
     \right\} 
 \nonumber \\
 & &
     - \frac{11}{6\eps} {\cal P}^{(0,1)}_{q \rightarrow q g} + \; c.c.,
\nonumber \\
\lefteqn{
{\cal P}^{(1,1)}_{(0,1)\; q \rightarrow q g}\left(T^{(1,1)}_a \right) 
 =  
 }
 \nonumber \\
 & &
 S_\eps^{-1}
 \frac{(4\pi)^\eps}{\Gamma(1-\eps)} 
      \left( \frac{s_{123}}{\mu^2} \right)^{-\eps}
  \left\{
        y^{-\eps}
          \left[ {\cal V}^{(0,1)}_{q \rightarrow q g} 
                +{\cal V}^{(0,1)}_{g \rightarrow g g}\left(\frac{(1-y)(1-z)}{1-z(1-y)}\right)
          \right]
 \right. \nonumber \\
 & & \left.
     +
      \left( 1-y \right)^{-\eps} \left( 1-z \right)^{-\eps}
          \left[ {\cal V}^{(0,1)}_{q \rightarrow q g} 
                +{\cal V}^{(0,1)}_{g \rightarrow g g}\left(\frac{y}{1-z(1-y)}\right)
          \right]
  \right\}
  \; {\cal P}^{(0,1)}_{q \rightarrow q g}.
\eq
The variables $y$ and $z$ are given by
\bq
y = \frac{s_{12}}{s_{123}}, & & z = \frac{s_{13}}{s_{13}+s_{23}}.
\eq
The functions $g_1$ and $g_2$ are given by
\bq
\lefteqn{
g_1(y,z) = }
 \nonumber \\
 & &
  - \frac{1}{\eps^2} 
 \left[ \Gamma(1+\eps) \Gamma(1-\eps) \left( \frac{z}{1-z} \right)^\eps + 1 
        - (1-y)^\eps z^\eps \; {}_2F_1\left( \eps, \eps, 1+\eps; (1-y)(1-z) \right) \right],
\nonumber \\
\lefteqn{
g_2 = \frac{1-\rho \eps}{2(1-\eps)(1-2\eps)}. }
\eq
The sum of the integrated dipoles reads
\bq
\label{ressingunresolved2}
{\cal V}^{(0,1)}_{q \rightarrow q g}+{\cal V}^{(0,1)}_{g \rightarrow g g}(r)
 & = & 
  \frac{\Gamma(-\eps) \Gamma(-2\eps) \Gamma(1-\eps)^2}{\Gamma(1-2\eps)\Gamma(1-3\eps)}
  \left\{ 4 + \frac{8 \eps}{1-3\eps} 
         - 2 \left(1-\rho \eps\right) \frac{\eps (1-\eps)}{(1-3\eps)(2-3\eps)} 
 \right. \nonumber \\
 & & \left.
         - \frac{4}{3} \frac{\eps(1-\eps)}{(1-3\eps)(2-3\eps)} r
  \right\}.
\eq
The amputated amplitudes are evaluated with the momenta
\bq
p_e & = & p_1 + p_2 - \frac{y}{1-y} p_3, 
\nonumber \\
p_s & = & \frac{1}{1-y} p_3.
\eq
For the numerical integration over the $(n+1)$-parton phase space we need the $\eps$-expansion of
the subtraction terms:
\bq
\lefteqn{
 \left\l A_{n}^{(0)} \right| {\cal P}^{(1,1)}(T) \left| A_{n}^{(0)} \right\r 
 = \left. A_2^{(0)} \right.^\ast A_2^{(0)} }
 \nonumber \\
 & &
 \times
\frac{2}{s_{12}} 
 \left\{
   \left[ \frac{2}{\eps^2} + \left( 3 - 2 L \right) \frac{1}{\eps} 
          + 2\;\mbox{Li}_2\left((1-y)(1-z)\right)
          + 2 \left( \ln(1-y)\right)^2 
 \right. \right. \nonumber \\
 & & \left. \left. 
          + 2 \ln(1-y) \ln(z) + 2 \ln(1-y) \ln(1-z) - 2 \ln(y) \ln(1-z)
          + 2 \ln(z) \ln(1-z) 
 \right. \right. \nonumber \\
 & & \left. \left. 
          - 2 \ln(y)\ln(1-y)
          - \frac{3}{2} \pi^2 
          - \frac{19}{6} \ln(1-z) - \frac{19}{6} \ln(1-y) - \frac{7}{2} \ln(y)
          + \frac{58}{3}
 \right. \right. \nonumber \\
 & & \left. \left. 
          + \frac{1}{3} \frac{(1-y)(1-z)}{1-z(1-y)} \left( \ln(y) - \ln(1-y) - \ln(1-z) \right)
          + L^2 - \frac{20}{3} L + \rho
   \right] 
 \right. \nonumber \\
 & & \left. \times 
   \left[ \frac{2}{1-z(1-y)} -2 + (1-y)(1-z) \right]
   +
   \left[ -\frac{2}{\eps} - 3 + 2L \right] \rho (1-y) (1-z)
   +
    1
 \right\} \nonumber \\
 & & + O(\eps).
\eq
Here, $L = \ln(s_{123}/\mu^2)$.
The $\eps$-expansion of the insertion of ${\cal P}^{(0,1)}$ into the interference
term with the loop amplitude $A_2^{(1)}$ is given by
\bq
\lefteqn{
         \left\l A_{n}^{(0)} \right| {\cal P}^{(0,1)}(T) \left| A_{n}^{(1)} \right\r 
       + \left\l A_{n}^{(1)} \right| {\cal P}^{(0,1)}(T) \left| A_{n}^{(0)} \right\r 
 = \left. A_2^{(0)} \right.^\ast A_2^{(0)}
 \times
 \frac{2}{s_{12}} 
 }
 \nonumber \\
 & &
 \times
 \left\{
   \left[ - \frac{2}{\eps^2} + \left( -3 + 2 L \right) \frac{1}{\eps} 
          - 8 + 3L - L^2 + \frac{7}{6} \pi^2 \right]
   \left[ \frac{2}{1-z(1-y)} -2 + (1-y)(1-z) \right]
 \right. \nonumber \\
 & & \left.
   +
   \left[ \frac{2}{\eps} + 3 - 2L \right] \rho (1-y) (1-z)
 \right\} + O(\eps).
\eq
We see that the poles in $\eps$ cancel in the sum of the two contributions, as it should be.

% section ------------------------------------------------------------------

\section{Cancellations}
\label{sectcanc}

In this section I discuss how singularities cancel between different terms
for the various singular limits.
The most intricate part is the one which is integrated over 
the $n+2$ parton phase space:
\bq
\label{nplus2finite}
 \int \left( {\cal O}_{n+2} \; d\sigma_{n+2}^{(0)} 
             - {\cal O}_{n+1} \circ d\alpha^{(0,1)}_{n+1}
             - {\cal O}_{n} \circ d\alpha^{(0,2)}_{n} 
      \right),
\eq
where
\bq
d \alpha^{(0,1)}_{n+1} =  
 \sum\limits_{T = T_a^{(1)}, ..., T_d^{(1)} } {\cal D}_{n+1}^{(0,1)}(T),
& &
d \alpha^{(0,2)}_{n} =  
 \sum\limits_{T = T_a^{(2)}, ..., T_h^{(2)} } {\cal D}_{n}^{(0,2)}(T).
\eq
Although individual terms in the integrand of eq. (\ref{nplus2finite}) are singular,
the singularities cancel in the sum and the integrand is
integrable. 
Therefore the integration can be performed by Monte Carlo methods.
It is instructive to see in detail, how the cancellations of singularities take place.
Critical region in phase space are regions where one or two particles are unresolved.
Double unresolved contributions can be divided into the following classes:
\begin{description}
\item{a)} Two pairs of separately collinear particles.
\item{b)} Three particles collinear.
\item{c)} Two particles collinear and a third soft particle.
\item{d)} Two soft particles.
\item{e)} Coplanar degeneracy.
\end{description}
The last item occurs only in the subtraction terms and 
is an artefact of our choice for the NLO subtraction terms.
It is included here for completeness.
The single unresolved contributions are divided into two classes:
\begin{description}
\item{f)} Two particles collinear.
\item{g)} One soft particle.
\end{description}
For the leading-colour contributions to $e^+ e^- \rightarrow q g g \bar{q}$
it is sufficient to check the following cases:
Particles $1$ and $2$ collinear and particles $3$ and $4$ collinear (case a);
particles $1$, $2$ and $3$ collinear (case b);
particles $1$ and $2$ collinear and particle $3$ soft (case c);
particles $2$ and $3$ soft (case d);
the sum of the four-momenta of particles $2$ and $3$ linear dependent with 
the four-momenta of particles $1$ and $4$ (case e);
particles $1$ and $2$ collinear (case f);
particles $2$ and $3$ collinear (also case f);
particle $2$ soft (case g).
All other cases are related by symmetry to the ones above.

All limits have been checked using the symbolic algebraic
manipulation program ``FORM'' \cite{Vermaseren:2000nd}.

\subsection{Double unresolved regions}

In the double unresolved cases, the singularities in the NLO subtraction
terms $d\alpha^{(0,1)}_{n+1}$ are cancelled by the subtraction terms
$d\alpha^{(0,2)}_{(0,1)\;n}$. In particular we always have
\bq
\label{cancdoubleunressubtr}
{\cal O}_{n+1} \; {\cal D}^{(0,1)}_{n+1}(T_a^{(1)})
-{\cal O}_{n} \; {\cal D}^{(0,2)}_{(0,1)\; n}(T_\topob^{(2)})
-{\cal O}_{n} \; {\cal D}^{(0,2)}_{(0,1)\; n}(T_\topoc^{(2)})
 & = &
\mbox{integrable},
\nonumber \\
{\cal O}_{n+1} \; {\cal D}^{(0,1)}_{n+1}(T_b^{(1)})
-{\cal O}_{n} \; {\cal D}^{(0,2)}_{(0,1)\; n}(T_\topod^{(2)})
-{\cal O}_{n} \; {\cal D}^{(0,2)}_{(0,1)\; n}(T_\topoh^{(2)})
 & = &
\mbox{integrable},
\nonumber \\
{\cal O}_{n+1} \; {\cal D}^{(0,1)}_{n+1}(T_c^{(1)})
-{\cal O}_{n} \; {\cal D}^{(0,2)}_{(0,1)\; n}(T_\topoe^{(2)})
-{\cal O}_{n} \; {\cal D}^{(0,2)}_{(0,1)\; n}(T_\topog^{(2)})
 & = &
\mbox{integrable},
\nonumber \\
{\cal O}_{n+1} \; {\cal D}^{(0,1)}_{n+1}(T_d^{(1)})
-{\cal O}_{n} \; {\cal D}^{(0,2)}_{(0,1)\; n}(T_\topoa^{(2)})
-{\cal O}_{n} \; {\cal D}^{(0,2)}_{(0,1)\; n}(T_\topof^{(2)})
 & = &
\mbox{integrable},
\eq
for all double unresolved cases.
Here, ``integrable'' stands for ``integrable over the double unresolved region''.
It remains to discuss the cancellations between $d\sigma^{(0)}_{n+2}$ and
$d\alpha^{(0,2)}_{(0,0)\;n}$ in the double unresolved regions.
It should be noted that in 
specific double unresolved limits some subtraction terms 
in eq. (\ref{cancdoubleunressubtr})
may be integrable by themselves. This will also be briefly 
discussed below case by case.

\subsubsection{Case $1$ and  $2$ collinear and $3$ and $4$ collinear.}

This case corresponds to the emission of two independent pairs 
of collinear particles ($p_1 || p_2$ and $p_3 || p_4$).
For each pair of collinear particles we write the corresponding momenta
according to eq. (\ref{collinearlimit}).
We are interested in contributions, which scale in the double
collinear limit as $|k_\perp|^{-4}$.
Cancellations of these terms occurs in the following combination:
\bq
{\cal O}_{n+2} \; d\sigma^{(0)}_{n+2}
-{\cal O}_{n} \; {\cal D}^{(0,2)}_{(0,0)\; n}(T_\topoa^{(2)})
-{\cal O}_{n} \; {\cal D}^{(0,2)}_{(0,0)\; n}(T_\topob^{(2)})
 & = &
\mbox{integrable}.
\eq
For completeness I also state the cancellations for terms related to
$d\alpha^{(0,1)}_{n+1}$:
\bq
{\cal O}_{n+1} \; {\cal D}^{(0,1)}_{n+1}(T_a^{(1)})
-{\cal O}_{n} \; {\cal D}^{(0,2)}_{(0,1)\; n}(T_\topob^{(2)})
 & = &
\mbox{integrable},
\nonumber \\
{\cal O}_{n+1} \; {\cal D}^{(0,1)}_{n+1}(T_d^{(1)})
-{\cal O}_{n} \; {\cal D}^{(0,2)}_{(0,1)\; n}(T_\topoa^{(2)})
 & = &
\mbox{integrable}.
\eq
All other subtraction terms are separately integrable in this limit.

\subsubsection{Case $1$, $2$ and $3$ collinear.}

This case corresponds to the triple collinear limit ($p_1||p_2||p_3$).
We parameterize the momenta of the three collinear particles
according to eq. (\ref{triplecollmom}).
We are interested in contributions, which scale in the double
collinear limit as $|k_\perp|^{-4}$.
Cancellations of these terms occurs in the following combination:
\bq
{\cal O}_{n+2} \; d\sigma^{(0)}_{n+2}
-{\cal O}_{n} \; {\cal D}^{(0,2)}_{(0,0)\; n}(T_\topoc^{(2)})
-{\cal O}_{n} \; {\cal D}^{(0,2)}_{(0,0)\; n}(T_\topod^{(2)})
-{\cal O}_{n} \; {\cal D}^{(0,2)}_{(0,0)\; n}(T_\topoe^{(2)})
 & = &
\mbox{integrable}.
 \nonumber \\
\eq
For completeness I also state the cancellations for terms related to
$d\alpha^{(0,1)}_{n+1}$:
\bq
{\cal O}_{n+1} \; {\cal D}^{(0,1)}_{n+1}(T_a^{(1)})
-{\cal O}_{n} \; {\cal D}^{(0,2)}_{(0,1)\; n}(T_\topoc^{(2)})
 & = &
\mbox{integrable},
\nonumber \\
{\cal O}_{n+1} \; {\cal D}^{(0,1)}_{n+1}(T_b^{(1)})
-{\cal O}_{n} \; {\cal D}^{(0,2)}_{(0,1)\; n}(T_\topod^{(2)})
 & = &
\mbox{integrable},
\nonumber \\
{\cal O}_{n+1} \; {\cal D}^{(0,1)}_{n+1}(T_c^{(1)})
-{\cal O}_{n} \; {\cal D}^{(0,2)}_{(0,1)\; n}(T_\topoe^{(2)})
 & = &
\mbox{integrable}.
\eq
All other subtraction terms are separately integrable in this limit.

\subsubsection{Case $1$ and $2$ collinear and $3$ soft.}

This case corresponds to the emission of a collinear pair ($p_1||p_2$) together
with a soft gluon ($p_3$).
In this limit, the invariants $s_{12}$, $s_{23}$, $s_{34}$ and $s_{123}$
become small.
We are interested in contributions, which have three of these four invariants 
in the denominator.
Cancellations of these terms occurs in the following combination:
\bq
{\cal O}_{n+2} \; d\sigma^{(0)}_{n+2}
- \sum\limits_{T = T_a^{(2)}, ..., T_g^{(2)} } 
 {\cal O}_{n} \; {\cal D}^{(0,2)}_{(0,0)\; n}(T)
 & = &
\mbox{integrable}.
\eq
For completeness I also state the cancellations for terms related to
$d\alpha^{(0,1)}_{n+1}$:
\bq
& &
{\cal O}_{n+1} \; {\cal D}^{(0,1)}_{n+1}(T_a^{(1)})
-{\cal O}_{n} \; {\cal D}^{(0,2)}_{(0,1)\; n}(T_\topob^{(2)})
-{\cal O}_{n} \; {\cal D}^{(0,2)}_{(0,1)\; n}(T_\topoc^{(2)})
 = 
\mbox{integrable},
\nonumber \\
& &
{\cal O}_{n+1} \; {\cal D}^{(0,1)}_{n+1}(T_c^{(1)})
-{\cal O}_{n} \; {\cal D}^{(0,2)}_{(0,1)\; n}(T_\topoe^{(2)})
 =
\mbox{integrable},
\nonumber \\
& &
{\cal O}_{n+1} \; {\cal D}^{(0,1)}_{n+1}(T_d^{(1)})
-{\cal O}_{n} \; {\cal D}^{(0,2)}_{(0,1)\; n}(T_\topoa^{(2)})
 = 
\mbox{integrable}.
\eq
All other subtraction terms are separately integrable in this limit.

\subsubsection{Case $2$ and $3$ soft.}

This case corresponds to the emission of two soft gluons 
(with momenta $p_2$ and $p_3$).
If we scale the momenta of two soft particles as
\bq
p_2 \rightarrow \lambda p_2, 
& &
p_3 \rightarrow \lambda p_3,
\eq
we are interested in contributions, which scale in the double
soft limit as $\lambda^{-4}$.
Cancellations of these terms occurs in the following combination:
\bq
{\cal O}_{n+2} \; d\sigma^{(0)}_{n+2}
- \sum\limits_{T = T_c^{(2)}, ..., T_h^{(2)} } 
 {\cal O}_{n} \; {\cal D}^{(0,2)}_{(0,0)\; n}(T)
 & = &
\mbox{integrable}.
\eq
For completeness I also state the cancellations for terms related to
$d\alpha^{(0,1)}_{n+1}$:
\bq
{\cal O}_{n+1} \; {\cal D}^{(0,1)}_{n+1}(T_b^{(1)})
-{\cal O}_{n} \; {\cal D}^{(0,2)}_{(0,1)\; n}(T_\topod^{(2)})
-{\cal O}_{n} \; {\cal D}^{(0,2)}_{(0,1)\; n}(T_\topoh^{(2)})
 & = &
\mbox{integrable},
\nonumber \\
{\cal O}_{n+1} \; {\cal D}^{(0,1)}_{n+1}(T_c^{(1)})
-{\cal O}_{n} \; {\cal D}^{(0,2)}_{(0,1)\; n}(T_\topoe^{(2)})
-{\cal O}_{n} \; {\cal D}^{(0,2)}_{(0,1)\; n}(T_\topog^{(2)})
 & = &
\mbox{integrable}.
\eq
All other subtraction terms are separately integrable in this limit.

\subsubsection{Coplanar case}

For completeness I briefly discuss singularities occuring in coplanar
configurations.
The matrix element squared $d\sigma^{(0)}_{n+2}$ is not singular in this limit.
This singularity occurs only in the subtraction terms.
An example is given by the subtraction term ${\cal D}^{(0,1)}_{n+1}(T^{(1)}_b)$.
This subtraction term has a singularity
in
\bq
\label{coplanarmom}
 2 p_4 p_e^{(1)} & = & \frac{s_{123}s_{234} - s_{23} s_{1234}}{s_{123}-s_{23}}.
\eq
The numerator is a Gram determinant and can also be written as
\bq
s_{123} s_{234} - s_{23} s_{1234} 
       & = & \l 1- | 2+3 | 4- \r \l 4- | 2+3 | 1- \r.
\eq
This expression vanishes, whenever the sum of the four-momenta $p_2+p_3$ becomes
linear dependent with $p_1$ and $p_4$.
Therefore, $2 p_4 p_e^{(1)}$ becomes small, whenever this condition is fullfilled and
whenever
$p_2+p_3$ is not linear dependent with $p_1$. (In the later case also the denominator
in eq. (\ref{coplanarmom}) vanishes.)
This type of singularity occurs only in the subtraction terms and the cancellation takes
place 
in the following combination:
\bq
{\cal O}_{n+1} \; {\cal D}^{(0,1)}_{n+1}(T_b^{(1)})
-{\cal O}_{n} \; {\cal D}^{(0,2)}_{(0,1)\; n}(T_\topoh^{(2)})
 & = &
\mbox{integrable}.
\eq

\subsection{Single unresolved regions}

The expression in eq. (\ref{cancdoubleunressubtr}) is integrated over the complete
$(n+2)$-parton phase space.
This implies in particular an integration over regions
which correspond to single unresolved configurations.
It is clear that in this case we cannot expect a cancellation of singularities
between terms,  which involve ${\cal O}_{n+2}$ and ${\cal O}_{n}$, e.g. similar
to that what we had in the double unresolved case.
In the single unresolved case the $(n+2)$-parton configuration used in the 
calculation of the observable ${\cal O}_{n+2}$ will go smoothly to a $(n+1)$-parton
configuration.
The numerical value for the observable ${\cal O}$ evaluated with this $(n+1)$-parton
configuration will however in general have no relationship with the numerical
value obtained from the evaluation with the $n$-parton configuration of the 
subtraction terms.
A similar argument applies to
the factorization of the matrix element.
Both effects spoil a cancellation of the singularities.
The cancellation of the singularities has to occur therefore either between
terms involving ${\cal O}_{n+2}$ and ${\cal O}_{n+1}$, or between
terms involving ${\cal O}_{n+1}$ and ${\cal O}_{n}$,
or between terms involving only ${\cal O}_{n}$.
In the following, 
``integrable'' stands for ``integrable over the single unresolved region''.

\subsubsection{Case $1$ and $2$ collinear.}

This case corresponds to the collinear limit $p_1||p_2$.
We parameterize the momenta of the two collinear particles
according to eq. (\ref{collinearlimit}).
We are interested in contributions, which scale in the 
collinear limit as $|k_\perp|^{-2}$.
A cancellations of these terms occurs in the following combination:
\bq
& &
{\cal O}_{n+2} \; d\sigma^{(0)}_{n+2}
-{\cal O}_{n+1} \; {\cal D}^{(0,1)}_{n+1}(T_a^{(1)})
 = 
\mbox{integrable}.
\eq
This is just the combination which already occurs at NLO.
At NNLO we also have to take into account the singularity in
${\cal D}^{(0,1)}_{n+1}(T_d^{(1)})$. This one is cancelled in the combination
\bq
{\cal O}_{n+1} \; {\cal D}^{(0,1)}_{n+1}(T_d^{(1)})
-{\cal O}_{n} \; {\cal D}^{(0,2)}_{(0,1)\; n}(T_\topoa^{(2)})
 = 
\mbox{integrable}.
\eq
Finally, there is a third equation, describing the cancellations among
the NNLO subtraction terms:
\bq
{\cal O}_{n} \; {\cal D}^{(0,2)}_{(0,0)\; n}(T_\topob^{(2)})
+{\cal O}_{n} \; {\cal D}^{(0,2)}_{(0,0)\; n}(T_\topoc^{(2)})
-{\cal O}_{n} \; {\cal D}^{(0,2)}_{(0,1)\; n}(T_\topob^{(2)})
-{\cal O}_{n} \; {\cal D}^{(0,2)}_{(0,1)\; n}(T_\topoc^{(2)})
 = 
\mbox{integrable}.
 \nonumber \\
\eq
All other subtraction terms are separately integrable in this limit.
Note that the pieces 
\bq
{\cal O}_{n} \; {\cal D}^{(0,2)}_{(0,0)\; n}(T_\topob^{(2)})
-{\cal O}_{n} \; {\cal D}^{(0,2)}_{(0,1)\; n}(T_\topob^{(2)})
\eq
and 
\bq
{\cal O}_{n} \; {\cal D}^{(0,2)}_{(0,0)\; n}(T_\topoc^{(2)})
-{\cal O}_{n} \; {\cal D}^{(0,2)}_{(0,1)\; n}(T_\topoc^{(2)})
\eq
are with our choice of subtraction terms 
in general {\bf not} integrable separately.
Only the sum of the two pieces is integrable.
In order for the cancellations to occur, the $n$-parton configuration
of topology $T_\topob^{(2)}$ must approach in the limit $s_{12} \rightarrow 0$
the same configuration as the $n$-parton configuration of topology
$T_\topoc^{(2)}$.
With a symmetric choice for the reconstruction functions in eq. (\ref{reconstructA}) and
eq. (\ref{reconstructC}) the $n$-parton configurations for topologies 
$T_\topob^{(2)}$ and $T_\topoc^{(2)}$ are identical for any value of $s_{12}$
and this requirement is fullfilled trivially.

\subsubsection{Case $2$ and $3$ collinear.}

Here we discuss the case, 
where particles $2$ and $3$ become collinear.
Again, we parameterize the momenta of the two collinear particles
according to eq. (\ref{collinearlimit}).
We are interested in contributions, which scale in the 
collinear limit as $|k_\perp|^{-2}$.
The NLO relation for the cancellation of singularities reads:
\bq
& &
{\cal O}_{n+2} \; d\sigma^{(0)}_{n+2}
-{\cal O}_{n+1} \; {\cal D}^{(0,1)}_{n+1}(T_b^{(1)})
-{\cal O}_{n+1} \; {\cal D}^{(0,1)}_{n+1}(T_c^{(1)})
 = 
\mbox{integrable}.
\eq
In addition, there is one relation among the subtraction terms
with $n$-parton configurations:
\bq
& &
{\cal O}_{n} \; {\cal D}^{(0,2)}_{(0,0)\; n}(T_\topod^{(2)})
+{\cal O}_{n} \; {\cal D}^{(0,2)}_{(0,0)\; n}(T_\topoh^{(2)})
-{\cal O}_{n} \; {\cal D}^{(0,2)}_{(0,1)\; n}(T_\topod^{(2)})
-{\cal O}_{n} \; {\cal D}^{(0,2)}_{(0,1)\; n}(T_\topoh^{(2)})
 \nonumber \\ & &
+{\cal O}_{n} \; {\cal D}^{(0,2)}_{(0,0)\; n}(T_\topoe^{(2)})
+{\cal O}_{n} \; {\cal D}^{(0,2)}_{(0,0)\; n}(T_\topog^{(2)})
-{\cal O}_{n} \; {\cal D}^{(0,2)}_{(0,1)\; n}(T_\topoe^{(2)})
-{\cal O}_{n} \; {\cal D}^{(0,2)}_{(0,1)\; n}(T_\topog^{(2)})
 \nonumber \\ & &
 \;\;\;
 = 
\mbox{integrable}.
\eq
All other subtraction terms are separately integrable in this limit.
Again we have to require that 
the $n$-parton configurations
of the topologies $T_\topod^{(2)}$, $T_\topoe^{(2)}$, $T_\topog^{(2)}$
and $T_\topoh^{(2)}$ all approach the same $n$-parton configuration
in the limit $s_{23} \rightarrow 0$.
With the choice for the reconstruction functions in eq. (\ref{reconstructD}) and
eq. (\ref{reconstructE}) this requirement is fullfilled.

\subsubsection{Case $2$ soft.}

This case corresponds to the emission of a single soft gluon (particle $2$).
If we scale the momentum of the soft gluon as
\bq
p_2 \rightarrow \lambda p_2, 
\eq
we are interested in contributions, which scale in the 
soft limit as $\lambda^{-2}$.
Cancellations of these terms occurs in the following combination:
\bq
& &
{\cal O}_{n+2} \; d\sigma^{(0)}_{n+2}
-{\cal O}_{n+1} \; {\cal D}^{(0,1)}_{n+1}(T_a^{(1)})
-{\cal O}_{n+1} \; {\cal D}^{(0,1)}_{n+1}(T_b^{(1)})
 = 
\mbox{integrable}.
\eq
This is the usual NLO relation.
In addition, we have
\bq
 & &
{\cal O}_{n} \; {\cal D}^{(0,2)}_{(0,0)\; n}(T_\topob^{(2)})
+{\cal O}_{n} \; {\cal D}^{(0,2)}_{(0,0)\; n}(T_\topoc^{(2)})
-{\cal O}_{n} \; {\cal D}^{(0,2)}_{(0,1)\; n}(T_\topob^{(2)})
-{\cal O}_{n} \; {\cal D}^{(0,2)}_{(0,1)\; n}(T_\topoc^{(2)})
 = 
\mbox{integrable},
 \nonumber \\
 & &
{\cal O}_{n} \; {\cal D}^{(0,2)}_{(0,0)\; n}(T_\topod^{(2)})
+{\cal O}_{n} \; {\cal D}^{(0,2)}_{(0,0)\; n}(T_\topoh^{(2)})
-{\cal O}_{n} \; {\cal D}^{(0,2)}_{(0,1)\; n}(T_\topod^{(2)})
-{\cal O}_{n} \; {\cal D}^{(0,2)}_{(0,1)\; n}(T_\topoh^{(2)})
 = 
\mbox{integrable}.
 \nonumber \\
\eq
All other subtraction terms are separately integrable in this limit.

\subsection{Cancellations for contributions with one virtual and one real unresolved parton}

The cancellation mechanism for contributions with one virtual and one
real unresolved parton are quite similar to those already encountered at NLO.
The only difference is the functional form of the subtraction terms 
(as given in eq. (\ref{oneloopsplitsubtrterm})), which approach smoothly the appropriate singular limit
(as given in eq. (\ref{oneloopcollinterference})).

% section ------------------------------------------------------------------

\section{Outlook and conclusions}
\label{sectconcl}

In this paper I considered the extension of the subtraction method to
next-to-next-to-leading order.
Such a method is needed to extend 
fully differential perturbative calculations
in quantum field theories from NLO to NNLO.
As a particular example I considered the leading-colour contributions
to $e^+ e^- \rightarrow \; \mbox{2 jets}$ and derived the subtraction
terms at NNLO.
The subtracted matrix elements can be integrated numerically
over the appropriate phase space.
The method presented here is general and not restricted to the example
of $e^+ e^- \rightarrow \; \mbox{2 jets}$.
Subtraction terms for other splittings (like $g \rightarrow g g g$)
and other kinematical configuartions (e.g. with partons in the initial state)
can be worked out along the same lines.

The subtraction terms still have to be integrated analytically 
over the unresolved
phase space.
With the advance of integration techniques witnessed in the past years,
this seems feasible. 
The completion of this program will open the door to fully differential
NNLO Monte Carlo programs.

\subsection*{Acknowledgements}
\label{sec:acknowledgements}

I would like to thank Sven Moch and Peter Uwer
for useful discussions and comments.

\end{document}